\documentclass[aip,jcp,reprint,amsmath,amssymb,nobalancelastpage]{revtex4-2}

\usepackage{graphicx}
\usepackage[suffix=]{epstopdf}
\usepackage{natmove}

\newcommand{\vv}[1]{\mathbf{#1}}

\begin{document}

\title{Approximation of anisotropic pair potentials using multivariate interpolation}

\author{Mohammadreza Fakhraei}
\affiliation{Department of Chemical Engineering, Auburn University, Auburn, AL 36849, United States}

\author{Chris A. Kieslich}
\email{kieslich@gatech.edu}
\affiliation{Department of Chemical Engineering, Auburn University, Auburn, AL 36849, United States}
\affiliation{Wallace H. Coulter Department of Biomedical Engineering, Georgia Institute of Technology, Atlanta, Georgia 30332, United States}

\author{Michael P. Howard}
\email{mphoward@auburn.edu}
\affiliation{Department of Chemical Engineering, Auburn University, Auburn, AL 36849, United States}

\begin{abstract}
The interaction between two particles with shape or interaction anisotropy can be modeled using a pairwise potential energy function that depends on their relative position and orientation; however, this function is often challenging to mathematically formulate. Data-driven approaches for approximating anisotropic pair potentials have gained significant interest due to their flexibility and generality but often require large sets of training data, potentially limiting their feasibility when training data is computationally demanding to collect. Here, we investigate the use of multivariate polynomial interpolation to approximate anisotropic pair potentials from a limited set of prescribed particle configurations. We consider both standard Chebyshev polynomial interpolation as well as mixed-basis polynomial interpolation that uses trigonometric polynomials for coordinates along which the pair potential is known to be periodic. We exploit mathematical reasoning and physical knowledge to refine the interpolation domain and to design our interpolants. We test our approach on two-dimensional and three-dimensional model anisotropic nanoparticles, finding satisfactory approximations can be constructed in all cases.
\end{abstract}

\maketitle

\section{Introduction}
Anisotropic particles are widely encountered as both synthetic \cite{boles:chemrev:2016,cai:chemsocrev:2021} and natural \cite{roth:biophysj:1996, hagan:advchemphys:2014, aumiller:chemsocrev:2018} materials, and the complex interactions between these particles is known to lead to interesting collective behaviors such as self-assembly \cite{solomon:colloidandinterface:2011, fejer:softmatter:2011, thorkelsson:nanotoday:2015}. For example, shape-anisotropic nanoparticles have been used to synthesize photonic crystals with desirable optical properties \cite{meseguer:colsurf:2005, kim:npasia:2011}, dense and exotic superlattices \cite{henzie:natmat:2012}, and porous supraparticles \cite{yektin:langmuir:2024}; anisotropic surface interactions have been designed to drive the formation of protein nanocages \cite{chen:advsci:2023}; and shape is believed to play a role in the aggregation of microplastics in the environment \cite{wang:jhazmat:2021, argun:softmatter:2023}. Computer simulations are useful tools for understanding and engineering these mesoscale processes, so there is interest to develop robust, efficient modeling capabilities for anisotropic particles.

An anisotropic particle can be represented as a single body with translational and rotational degrees of freedom \cite{allen:oxford:2017}, and a pairwise potential energy function can be used to model how the effective interaction between two anisotropic particles depends on their relative position and orientation. Examples of such energy functions include the Kern--Frenkel potential for patchy nanoparticles \cite{kern:jchemphys:2003}, the Gay--Berne potential for ellipsoids \cite{berardi:chemphyslett:1998}, and an anisotropic generalization of the Weeks--Chandler--Andersen repulsive potential \cite{weeks:jchemphys:1971} to hard polyhedra \cite{ramasubramani:jchemphys:2020}. These pair potentials have functional forms that are relatively straightforward to parametrize and implement in simulations but only represent a limited set of anisotropic particles. Additional types of anisotropic particles can be described as composites of smaller constituent particles that interact isotropically \cite{nguyen:cpc:2019}. An effective anisotropic interaction arises from the placement of the constituent particles and the nature of their interactions with each other. This approach offers the flexibility to represent many types of anisotropic particles \cite{wani:softmatter:2024, argun:softmatter:2023}; however, such models can be cumbersome to construct and their computational cost grows with the number of constituent particles required.

Recently, there has been significant interest in using data-driven methods to approximate anisotropic pair potentials with unknown functional dependence on the relative position and orientation of the particles. An anisotropic potential modeled as a multivariate series of cubic B-splines and cosines was fit to measured forces and torques \cite{nguyen:jchemphys:2022} using a generalization of the multiscale coarse-graining method \cite{noid:jchemphys:2008_1, noid:jchemphys:2008_2}. Linear regression of a downselected pool of particle-centered descriptors \cite{campos-villalobos:jchemphys:2022,campos-villalobos:npjcompmat:2024} and density expansions \cite{lin:jchemphys:2024} has been used to approximate anisotropic pair potentials without prescribing the functional form. Neural-network potentials were also used to model anisotropic interactions of benzene and sexithiophene \cite{wilson:jchemphys:2023} as well as shape-anisotropic nanoparticles \cite{argun:jchemphys:2024} by force and torque matching. These data-driven models have shown promising accuracy but training them can require large amounts of data that may not always be feasible to collect, e.g., if the underlying model is costly to simulate.

In this work, we explore the possibility of using multivariate interpolation to approximate anisotropic pair potentials with limited data. Specifically, we consider an $N$-term approximation $\hat u$ of the true pair potential $u$ that is a function of the relative position $\vv{r}$ and relative orientation $\boldsymbol{\Omega}$ of two anisotropic particles,
\begin{equation}
\hat u(\boldsymbol{\vv{r}, \boldsymbol{\Omega}}) = \sum_{n=0}^{N-1} c_n \psi_n(\boldsymbol{\vv{r}, \boldsymbol{\Omega}}),
\label{eq:surrogate}
\end{equation}
where $c_n$ and $\psi_n$ are the $n$-th coefficient and basis function, respectively. The set of coefficients $\{c_n\}$ can be uniquely determined from a system of linear equations equating $\hat u$ and $u$ at only $N$ points where $u$ has been sampled. Different interpolation schemes typically consist of both a set of basis functions and a set of sample points that are connected to each other in some way. For example, piecewise linear interpolation is usually performed with uniformly spaced sample points that are colocated with the boundaries of the locally linear pieces, whereas Chebyshev polynomial interpolation is typically performed using sample points placed at the roots or extrema of the Chebyshev polynomials of the first kind that also serve as global basis functions. This relationship between basis functions and sample points for interpolation schemes differs from black-box machine-learning approaches, for which the sample points are usually only expected to cover the input space, and it can help produce good approximations with relatively small amounts of data. For example, Chebyshev polynomial interpolation using the Chebyshev extrema as sample points tends to minimize oscillations near the edges of the approximation domain \cite{press:cambridge:2007, trefethen:socindmath:2019}.

Polynomials with analogous form to eq \eqref{eq:surrogate} have been successfully used to construct classical potential energy functions. The Chebyshev Interaction Model for Efficient Simulation (ChIMES)\cite{lindsey:jchemtheorycomp:2017, lindsey:jchemtheorycomp:2019, lindsey:jchemphys:2020, lindsey:jchemphys:2023} approximates the many-body potential of mean force (up to four-body terms) using a multivariate polynomial in the distances between bodies whose basis functions are products of Chebyshev polynomials of the first kind. ChIMES has been demonstrated to accurately model molten carbon \cite{lindsey:jchemtheorycomp:2017}, water \cite{lindsey:jchemtheorycomp:2019}, carbon condensation \cite{lindsey:jchemphys:2020}, and nitrogen under shock compression\cite{lindsey:jchemphys:2023}. In ChIMES, the coefficients of the basis functions are fit to match the forces observed in a quantum-mechanical simulation using least-squares regression rather than interpolate the energy. Nance and co-workers have used Langrange polynomials with sparsely sampled Chebyshev extrema to interpolate an \textit{ab initio} potential energy surface and reaction path for the isomerization of 2-butene \cite{nance:jchemtheroycomp:2014,nance:jscicomp:2015}. Additionally, Morrow et al.~used a combination of Lagrange polynomials and trigonometric polynomials with sparse sampling to model a potential energy surface for azomethane \cite{morrow:jchemtheorycomp:2021}. Based on physical knowledge, they used Lagrange polynomials as basis functions for nonperiodic coordinates (e.g., bond lengths and bond angles) and trigonometric polynomials as basis functions for periodic coordinates (e.g., torsional angles). Nguyen and Huang  also used a mix of basis functions for force and torque matching of anisotropic potentials \cite{nguyen:jchemphys:2022}; in their work, the radial-distance dependence was captured by B-spline functions, while all other angular dependences were expressed using cosines.

Building on this prior work, we develop and assess a framework for using multivariate interpolation to approximate anisotropic pair potentials. Anisotropic interactions present several challenges for interpolation that must be addressed. First, the prescribed sample points may contain overlapping configurations with large repulsion (large or divergent energies) that are challenging to accurately approximate using polynomials, but these configurations may occur at varying center-to-center separations of the particles due to their anisotropy. Second, physical symmetries of the shapes may introduce aliasing effects or sampling inefficiencies. Finally, the interaction energy may vary much more significantly with respect to some coordinates than others. We address these challenges by developing suitable coordinate transformations to remove unfavorable configurations from the approximation domain, leveraging physical symmetries to reduce the approximation domain, and choosing appropriate basis functions and degrees of approximation for our transformed coordinates based on both physical and mathematical considerations. Overall, we find that our approach produces quite satisfactory approximations using relatively few terms (and corresponding sample points) in eq \eqref{eq:surrogate}.

The rest of this article is organized as follows. We first describe the model two- and three-dimensional anisotropic nanoparticles, comprised of Lennard-Jones particles, that we considered as test problems. We then develop our framework for interpolating anisotropic pair potentials using a two-dimensional rod nanoparticle as a motivating example. We apply our framework to approximate anisotropic pair potentials for all our two- and three-dimensional nanoparticles and assess its accuracy. We conclude with a summary of our key results and some remarks on the framework.

\section{Model anisotropic nanoparticles}
\begin{figure*}
    \includegraphics{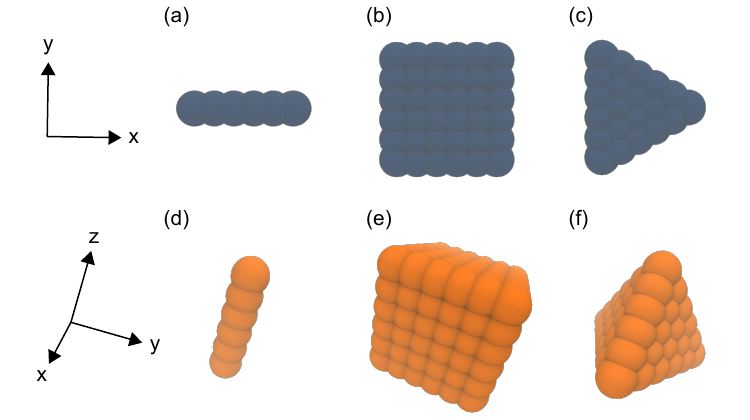}
    \caption{Two-dimensional nanoparticles [(a) rod, (b) square, and (c) triangle] and three-dimensional nanoparticles [(d) rod, (e) cube, and (f) tetrahedron] in their initial orientations. Images rendered using VMD 1.9.4 \cite{humphrey:jmolgrp:1996}.}
    \label{fig:shapes}
\end{figure*}

We used three two-dimensional (rod, square, triangle) and three three-dimensional (rod, cube, tetrahedron) anisotropic nanoparticles as representative test problems for our framework (Figure \ref{fig:shapes}). All nanoparticles were modeled using spherical constituent particles with nominal diameter $\sigma$ that were regularly arranged inside the nanoparticle volume and separated from their nearest neighbors by $2\sigma/3$. The constituent-particle discretization was also similar for all nanoparticles (6 constituent particles per axis or edge). The initial position and orientation of a nanoparticle was defined such that its center of mass was at the origin and its moment of inertia tensor was diagonal (i.e., its principal axes were aligned with the Cartesian axes). The two-dimensional nanoparticles translated in $x$ and $y$, and they rotated about the $z$-axis by angle $\alpha$. The three-dimensional nanoparticles translated in $x$, $y$, and $z$, and they rotated using an intrinsic (body-attached) Euler angle sequence in which the nanoparticle was first rotated about its $z$-axis by $\alpha$, then about its $x$-axis by $\beta$, and last about its $z$-axis again by $\gamma$.

The total energy between two nanoparticles was computed as a pairwise sum of interactions between constituent particles in different nanoparticles. The constituent particles interacted through a perturbed Lennard-Jones potential,\cite{weeks:jchemphys:1971}
\begin{equation}
u(r) = \begin{cases}
    u_{\rm LJ}(r) + (1-\lambda)\varepsilon,  & r \leq 2^{1/6}\sigma \\
    \lambda u_{\rm LJ}(r), & r > 2^{1/6}\sigma
\end{cases}, \label{eq:true_potential}
\end{equation}
where $u_{\rm LJ}$ is the standard Lennard-Jones potential,
\begin{equation}
u_{\rm LJ}(r) = 4 \varepsilon \left[\left(\frac{\sigma}{r}\right)^{12} - \left(\frac{\sigma}{r}\right)^6 \right],
\label{eq:lj}
\end{equation}
$\varepsilon$ is the energy scale, $\lambda$ modulates the strength of the attraction, and $r$ is the distance between the centers of the constituent particles. The constituent-particle interaction was truncated and shifted to zero at $3\,\sigma$ for computational convenience. To help reduce the dependence of our model on the constituent-particle discretization across different nanoparticle shapes, we numerically computed $\lambda$ to make the interaction energy $-5\varepsilon$ for the most attractive configuration for each nanoparticle.  We first minimized $u$ with respect to $\vv{r}$ and $\boldsymbol{\Omega}$ when $\lambda = 1$, starting from a edge-to-edge or face-to-face configuration. We then performed a bisection search for $0 \le \lambda \le 1$ to solve for $\lambda$ such that $u = -5\varepsilon$ for this configuration.

\section{Framework}
\subsection{Multivariate interpolation}
We considered multivariate interpolation schemes using continuous, differentiable basis functions on the domain of $\hat u$ because the gradient of $\hat u$ is required for some energy minimization methods and to evaluate forces and torques for molecular dynamics simulations \cite{allen:oxford:2017}. The multivariate basis functions and sample points were constructed using a tensor product that forms all possible combinations of univariate basis functions and sample points selected for each coordinate. Specifically, we used univariate interpolation schemes based on Chebyshev polynomials and trigonometric polynomials. For the Chebyshev polynomial scheme, the univariate basis functions were the first $P+1$ Chebyshev polynomials of the first kind $T_n(x)$ for $n = \{0, ..., P\}$, which have the domain $-1 \le x \le 1$. The corresponding sample points were the extrema of the highest degree polynomial $T_P$, which occur at $\cos(\pi m / P)$ for $m = \{0, ..., P\}$ when $P \ge 1$ and nominally at 0 for $P = 0$. For the trigonometric scheme, the univariate basis functions were the $P+1$ complex exponential functions $\phi_n(x) = e^{i n x}$ for $n = \{-P/2, ..., P/2\}$, which have the domain $0 \le x < 2\pi$; here, $P$ is required to be even. The corresponding sample points were uniformly spaced in the domain of the basis function, at $2 \pi m / (P + 1)$ for $m = \{0, ..., P\}$. In both cases, the domain of the coordinate being interpolated was linearly transformed to the domain of the basis function. We anticipated that the Chebyshev polynomials might be better basis functions for nonperiodic coordinates, while the trigonometric polynomials might be better basis functions for periodic coordinates. We will discuss this more later.

\begin{figure}
    \includegraphics{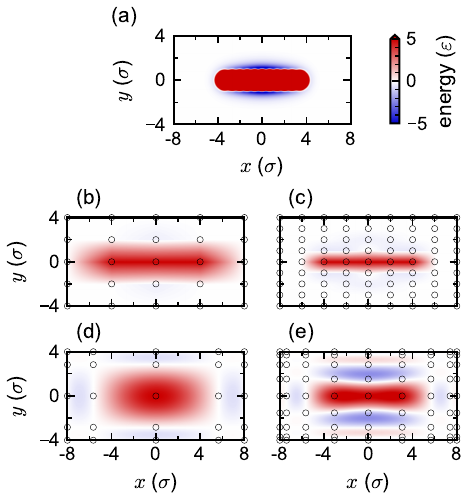}
    \caption{Energy of a pair of two-dimensional rods fixed in their initial orientation as a function of relative Cartesian position: (a) true, (b--c) piecewise linear interpolation using Cartesian coordinates, and (d--e) Chebyshev polynomial interpolation using Cartesian coordinates. In (b) and (d), 5 sample points were used per coordinate, while in (c) and (e), 9 sample points were used per coordinate.}
    \label{fig:rod_cart}
\end{figure}

We initially attempted to interpolate $u(x,y)$ for a pair of two-dimensional rods fixed in their initial orientation as a function of the Cartesian position $(x, y)$ of one relative to the other (Figure \ref{fig:rod_cart}). The domain of the interpolant  $\hat u$ was $-8\,\sigma \le x \le 8\,\sigma$ and $-4\,\sigma \le y \le 4\,\sigma$, and we considered 5 sample points ($P = 4$) for both coordinates. To establish a reference point for our chosen polynomial interpolation schemes, we also constructed a multivariate linear piecewise interpolant on an equivalent-size uniform grid of sample points. The linear piecewise interpolant is simpler to evaluate than the  polynomial interpolant but is not continuously differentiable. For some sample points, $u$ diverged or was very large, so we capped the energy to be less than $5\,\varepsilon$ as a typical ``large'' value that would not be expected to be frequently observed in a simulation. Both approximations were poor compared to the true pair potential [Figure \ref{fig:rod_cart}(a)]. The piecewise linear interpolant had limited resolution of the repulsion and did not capture the deep attraction between side-by-side rods [Figure \ref{fig:rod_cart}(b)]. The Chebyshev polynomial interpolant did worse at capturing the repulsion and also exhibited unphysical attraction at large separations [Figure \ref{fig:rod_cart}(d)]. Increasing the resolution of the interpolation to use 9 sample points ($P = 8$) per coordinate did not resolve these issues [Figures \ref{fig:rod_cart}(c) and \ref{fig:rod_cart}(e)].

This result highlights one of the key challenges for interpolating anisotropic pair potentials. By construction, the multivariate sample points form a grid that spans the entire domain; however, some of these sample points include energetically-unfavorable configurations that decrease the accuracy of the interpolation for other configurations and, practically, may not be as important to faithfully resolve because they are less likely to be observed. Machine-learning-based approximations are less susceptible to this complication because their sample space is not as rigidly prescribed and configurations can be straightforwardly excluded from the training data. Indeed, constant-temperature molecular simulations have often be used to collect the training data for such models, so these high-energy configurations are naturally excluded by the simulation. \cite{argun:jchemphys:2024, campos-villalobos:jchemphys:2022, campos-villalobos:npjcompmat:2024, wilson:jchemphys:2023, nguyen:jchemphys:2022} A different strategy is required to remove such configurations from interpolative approximations.

\subsection{Variable transformation}
\begin{figure*}
    \includegraphics{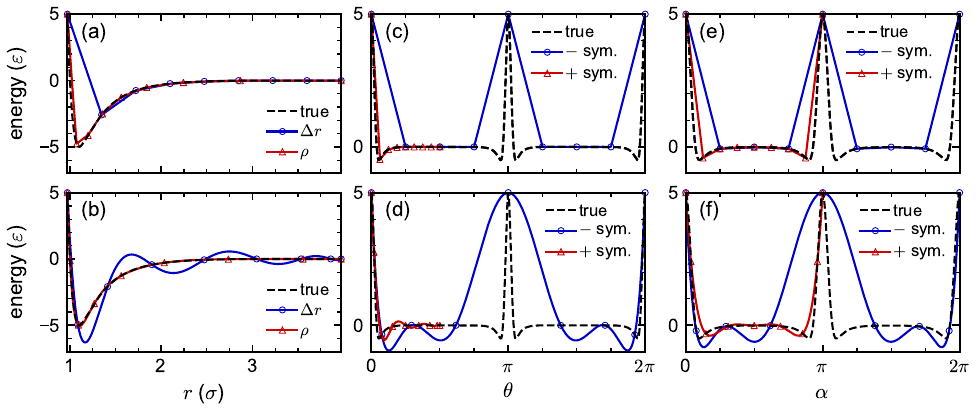}
    \caption{Energy for a pair of two-dimensional rods as a function of (a--b) $r$ at fixed $\theta$ and $\alpha$, (c--d) $\theta$ at fixed $r$ and $\alpha$, and (e--f) $\alpha$ at fixed $r$ and $\theta$ for the configurations described in the text. The true potential energy is compared to one-dimensional linear piecewise interpolants (top row) or Chebyshev polynomial interpolants (bottom row). In (a--b), the interpolants were constructed using either $\Delta r$ or $\rho$ as the transformed distance coordinate, while in (c--f), the interpolants were constructed either without ($-$) or with ($+$) symmetry (sym.) All interpolants used 9 sample points.}
    \label{fig:rod_1dinterp}
\end{figure*}

For many types of particles, the pair potential is expected to rapidly increase to enforce a strong, short-ranged repulsion when the particles approach a configuration in which their volumes overlap. For example, the isotropic Lennard-Jones potential [eq \eqref{eq:lj}] increases drastically when $r < \sigma$ (two spheres overlap), so highly energetically unfavorable configurations can be excluded by setting a lower bound on $r$. For anisotropic pair potentials, though, an analogous lower bound on the particles' separation is more complex to prescribe because it can depend on both the relative position $\vv{r}$ and orientation $\boldsymbol{\Omega}$ of the particles. For example, our two-dimensional rods can only approach to a center-to-center distance of $4.3\,\sigma$ in an end-to-end configuration but to $1\,\sigma$ in a side-to-side configuration if their volume is nominally that of a spherocylinder.

To address this issue, we first represented the relative particle position $\vv{r}$ in spherical coordinates $(r,\theta,\phi)$, where $r$ is the center-to-center distance, $0 \le \theta < 2\pi$ is the azimuthal angle, and $0 \le \phi \le \pi$ is the polar angle. Both angles used the standard convention relative to the principal axes of the reference particle chosen as the origin. We then defined $r_0(\theta,\phi,\boldsymbol{\Omega})$ as a position- and orientation-dependent lower bound on $r$. Different criteria for $r_0$ could be adopted; in this work, we defined $r_0$ as the largest $r$ for which $u$ exceeded a threshold $5\,\varepsilon$ because the probability of observing particles in such a configuration with larger energy is unlikely if $\varepsilon$ is comparable to thermal energy. For our nanoparticles, we solved for $r_0$ numerically: for a given $(\theta, \phi, \boldsymbol{\Omega})$, we performed a decreasing linear search to find the first $r$ where $u > 5\, \varepsilon$ using a small step size ($0.1\,\sigma$) and starting from an initial guess at which $u < 5\varepsilon$ for all configurations, then we refined the resulting bracket using bisection search.

An upper bound on $r$ is also typically adopted to reduce the number of pair interactions that must be evaluated in a simulation \cite{allen:oxford:2017}. Inspired by core-shifted forms of eq \eqref{eq:lj} and the Gay--Berne model for ellipsoids, we assumed that anisotropic interactions resulting from short-ranged attractions would sufficiently decay after the center-to-center distance was increased by a nominal cutoff distance $r_{\rm c}$, making the approximation domain $r_0 \le r \le r_0 + r_{\rm c}$. We used $r_{\rm c} = 3\,\sigma$ for all nanoparticles in this work.

In our multivariate interpolation schemes, the domain of each coordinate must be specified independently of the others, so these bounds on $r$ are not immediately suitable for this purpose. We first considered a variable redefinition $\Delta r = r - r_0$, which is bounded $0 \le \Delta r \le r_{\rm c}$ and so is suitable for interpolation. To test the effectiveness of this redefinition, we constructed a one-dimensional interpolant $\hat u(\Delta r)$ for the side-to-side configuration of our two-dimensional rod ($\theta = \pi/2$, $\alpha = 0$) using either piecewise linear interpolation [Figure \ref{fig:rod_1dinterp}(a)] or Chebyshev polynomial interpolation [Figure \ref{fig:rod_1dinterp}(b)] with 9 sample points ($P=8$). While we observed significant improvements compared to equivalent traces of Figure \ref{fig:rod_cart} (interpolation using Cartesian coordinates), there were still visible errors in the approximation. The piecewise linear interpolant poorly approximated the attractive minimum due to the lack of samples in that region, while the Chebyshev polynomial interpolant  exhibited oscillatory behavior that we attribute to attempting to represent a diverging function using a polynomial. 

Based on these shortcomings, we considered an alternative variable redefinition that leverages physical knowledge. Since many pair potentials are known to depend on powers of $1/r$ (rather than $r$), we defined
\begin{equation}
\rho = \frac{1/r - 1/r_0}{1/(r_0+r_{\rm c}) - 1/r_0},
\label{eq:rho}
\end{equation}
where $0 \le \rho \le 1$. Both the piecewise linear interpolant and Chebyshev polynomial interpolant based on $\rho$ improved substantially compared to the same based on $\Delta r$ using the same number of points. The linear piecewise interpolant still had some small inaccuracies in approximating the attractive minimum, but the Chebyshev polynomial interpolant was in excellent agreement with the true energy across the entire domain. We believe that the transformation to $\rho$ helped both by placing more sample points at shorter distances where the energy changed more rapidly, as well as by effectively changing the functional form of the interpolation.

\subsection{Symmetry}
\begin{table*}
    \caption{Upper bounds of angles sampled for interpolation, solid angle represented, and corresponding reduction relative to the total solid angle ($4\pi^2$ for two-dimensional nanoparticles, $32\pi^3$ for three-dimensional nanoparticles).}
    \begin{tabular}{cccccccc}
    nanoparticle & $\theta$ & $\phi$ & $\alpha$ & $\beta$ & $\gamma$ & solid angle & reduction\\
    \hline
    rod (2D) & $\pi/2$ &  & $\pi$ & & & $\pi^2/4$ & 8\\
    square & $\pi/4$ &  & $\pi/2$ & & & $\pi^2/8$ & 32 \\
    triangle & $\pi/3$ &  & $2\pi/3$ & & & $2\pi^2/9$ & 18  \\
    rod (3D) &  & $\pi/2$ & $2\pi$ & $\pi/2$ & & $2\pi$ & $\approx 158$  \\
    cube & $\pi/4$ & $\pi/2$ & $2\pi$ & $\cos^{-1}(1/\sqrt{3})$ & $\pi/2$ & $(3-\sqrt{3})\pi^3/12$ & $\approx 303$\\
    tetrahedron & $2\pi/3$ & $\pi$ & $2\pi$ & $\pi$ & $2\pi/3$ & $32\pi^3/9$ & 9
    \end{tabular}
    \label{tab:angles}
\end{table*}
Having accurately approximated the energy $u$ with respect to only the center-to-center distance $r$, we then considered schemes for each of the other translational coordinates as well as the rotational coordinates. We constructed a one-dimensional interpolant $\hat u(\theta)$ for two-dimensional rods fixed in their initial orientation ($\alpha = 0$) and another interpolant $\hat u(\alpha)$ for two-dimensional rods initially in an end-to-end configuration ($\theta = 0$) using the same number of points as for $\hat u(\rho)$. In both cases, $r$ was fixed at the maximum value of $r_0$ as $\theta$ or $\alpha$ was varied. Unfortunately, both the piecewise linear interpolants [Figures \ref{fig:rod_1dinterp}(c) and \ref{fig:rod_1dinterp}(e)] and the Chebyshev polynomial interpolants [Figures \ref{fig:rod_1dinterp}(d) and \ref{fig:rod_1dinterp}(f)] were highly inaccurate. One cause of this approximation error is that $u$ quickly varies when the rod passes through configurations close to overlap, similarly to how $u$ increases when $r$ decreases. However, unlike in the case of $r$, there was not an obvious variable transformation that could be used to remove such configurations because some of them occurred in the middle of the approximation domain.

We additionally noted that $u$ was periodic with respect to $\theta$ and $\alpha$ due to physical symmetries of the rod. The other nanoparticles we considered also have symmetries, and symmetry is often present to some extent in many anisotropic particles that are of scientific interest \cite{pearce:nature:2021, damasceno:science:2012}. We hypothesized that accounting for symmetry (periodicity) when defining our approximation domain would improve the accuracy of $\hat u$. Symmetry arguments were used in ref \citenum{argun:jchemphys:2024} to reduce the input space to the neural-network anisotropic pair potential, with mixed success reported.

For the translational spherical coordinates $(\theta,\phi)$, we considered symmetries of the reference particle (serving as the origin for the interaction) that could be expressed as proper rotations that preserve the orientation of the coordinate system. We then rotated the second particle around the reference particle to reduce $\theta$ and $\phi$ to their smallest equivalent values. For example, a pair of two-dimensional rods with relative position $\theta \ge \pi$ can be rotated about the $z$-axis of the reference particle by $\pi$. The reference particle is in an equivalent configuration after this rotation, but the second particle now has $\theta < \pi$. Further, if $\theta > \pi / 2$ after this transformation, the rods can be rotated again about the new $y$-axis of the reference particle by $\pi$. This transformation again leaves the reference particle in an equivalent configuration, but now $0 \le \theta \le \pi /2$.

The orientation coordinates represented by Euler angles ($0 \le \alpha < 2\pi$, $0 \le \beta \le \pi$, and $0 \le \gamma < 2\pi$) can also be reduced using symmetry. A procedure for doing so has been used previously in the context of crystallography to standardize orientation descriptors \cite{nolze:cryst:2015}. For our Euler angle convention, the domain of $\gamma$ for three-dimensional particles ($\alpha$ for two-dimensional particles) can be reduced by the degree of rotational symmetry about the principal axis aligned with the $z$-axis. For example, the two-dimensional rod can be rotated by $\pi$ to an equivalent configuration, so $0 \le \alpha < \pi$ is sufficient to represent its orientations. We recommend that the principal axis with the highest degree of rotational symmetry should be aligned in this way. Additionally, for three-dimensional particles, the domain of $\beta$ can be reduced to $0 \le \beta \le \pi/2$ if the principal axis aligned with the $x$-axis has two-fold rotational symmetry by remapping $\alpha$ to $\pi + \alpha$, $\beta$ to $\pi - \beta$, and $\gamma$ to $2\pi-\gamma$ \cite{nolze:cryst:2015}. We applied these strategies for all our nanoparticles, reducing the solid angle represented by a factor between $8$ and $303$  (Table \ref{tab:angles}). Details of the rotations needed to reduce the translational coordinates for each nanoparticle are described in the Supporting Information.

To test how much leveraging symmetry might improve approximation accuracy, we constructed new interpolants $\hat u(\theta)$ and $\hat u(\alpha)$ for the two-dimensional rods using the same number of points in the reduced domains that account for symmetry. Both the piecewise linear interpolants [Figures \ref{fig:rod_1dinterp}(c) and \ref{fig:rod_1dinterp}(e)] and the Chebyshev polynomial interpolants [Figures \ref{fig:rod_1dinterp}(d) and \ref{fig:rod_1dinterp}(f)] that used symmetry agreed much better with the true energy than those that did not. By removing periodicities due to known symmetry, the sampling density of unique configuration space effectively increases, which is important when approximating using limited data.

\subsection{Selecting univariate approximations}
\begin{figure}
    \includegraphics{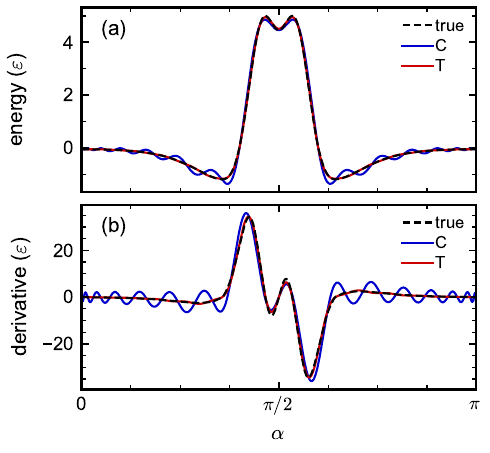}
    \caption{(a) Energy for a pair of two-dimensional rods as a function of $\alpha$ with $\theta = \pi/2$ and $r$ fixed at the maximum value of $r_0$ as $\alpha$ varies. The true energy is compared to a Chebyshev polynomial interpolant (C) using 33 sample points ($P=32$) and a trigonometric polynomial interpolant (T) using 27 sample points ($P=26$). (b) The first derivative of the same with respect to $\alpha$.}
    \label{fig:derivative}
\end{figure}
In addition to physical symmetries of specific anisotropic particles, some of the coordinates chosen for our approximation have inherent periodicity ($\theta$, $\alpha$, and $\gamma$). Application of symmetry to reduce their domains may remove some of these periodicities but some may remain; indeed, $\alpha$ and $\gamma$ for three-dimensional particles and $\alpha$ for two-dimensional particles are always periodic using our procedure. We typically expect that not only the energy but also its partial derivatives with respect to periodic coordinates should be periodic. The sample points for the Chebyshev polymomials naturally enforce periodicity of the interpolant, but its first derivative is not guaranteed to be periodic. Trigonometric polynomials, on the other hand, are inherently periodic functions that also have periodic derivatives; they work well for approximating periodic data but can struggle to represent nonperiodic data. Hence, a ``mixed'' basis \cite{morrow:jphyschemb:2019, morrow:jscicomp:2020, morrow:jchemtheorycomp:2021}, formed from a tensor product of Chebyshev polynomials for nonperiodic coordinates and trigonometric polynomials for periodic coordinates, may be advantageous for multivariate interpolation of anisotropic pair potentials.

In support of this notion, we compared the use of Chebyshev polynomials and trigonometric polynomials to interpolate $\hat u(\alpha)$ for two-dimensional rods with $\theta=\pi/2$ and $r$ fixed at the maximum $r_0$ as $\alpha$ varied (Figure \ref{fig:derivative}). To ensure a reasonably faithful approximation of the energy, we used 33 points for the Chebyshev polynomial interpolant and $27$ points for the trigonometric polynomial interpolant. (The reasons for selecting these particular numbers of points are described below.) The trigonometric polynomial interpolant, despite having slightly fewer terms and sample points, produced a better approximation of $\hat u$ than the Chebyshev polynomial interpolant, which showed unphysical oscillations. We also computed ${\rm d}\hat u/{\rm d}\alpha$, and these differences were more exaggerated. We noted that ${\rm d}\hat u/{\rm d}\alpha$ was also discontinuous across the periodic boundary for the Chebyshev polynomial interpolant, although the true energy and the trigonometric polynomial interpolant were continuous. This example highlights the potential benefits of selecting trigonometric polynomials for periodic coordinates.

After selecting the type of basis function for a given coordinate, the degree of the univariate polynomial must also be chosen. We anticipated that some coordinates may be more challenging to interpolate than others, requiring a higher degree polynomial and more sample points. Even for interpolating a single coordinate, some configurations may be more challenging than others, as can be seen for the side-to-side, side-to-end, and end-to-end configurations of our two-dimensional rods (Figure S1). To efficiently refine the approximation of a given coordinate, we considered only univariate polynomial degrees that produced nested sample points, i.e., all sample points continued to be used as the degree increased. Nested points of a nominal nonnegative integer ``level'' $l$ were generated by choosing $P = 2^\ell$ when $\ell \geq 1$ and $P = 0$ when $\ell = 0$ for Chebyshev polynomials and $P = 3^\ell-1$ for trigonometric polynomials. We considered $2^\ell+1$ nested sample points for the corresponding linear piecewise linear interpolant.

\section{Results and Discussion}
We tested our framework on the model two-dimensional and three-dimensional nanoparticles we selected as test problems (Figure \ref{fig:shapes}). We constructed a linear piecewise interpolant, a Chebyshev polynomial interpolant, and a mixed-basis polynomial interpolant that used Chebyshev polynomials for the nonperiodic coordinates and trigonometric polynomials for the periodic coordinates. All interpolants used $\rho$ as the transformed coordinate for the center-to-center distance, and Table \ref{tab:angles} lists the angles with corresponding upper bounds for each nanoparticle after application of symmetry. The periodic coordinates were $\alpha$ and $\gamma$.

To interrogate how much data was required to accurately approximate the energy $u$, we exhaustively searched combinations of univariate approximations of different degrees that produced a multivariate interpolant $\hat u$ with fewer than a threshold number of sample points [$N$ in eq \eqref{eq:surrogate}]. In this discussion, we emphasize that the number of sample points also describes the number of basis functions included in the polynomial interpolants, and interpolants with more sample points are expected to be more accurate because they have sampled more configurations and have more flexibility in their functional forms. Based on Figure S1 for the two-dimensional rod, we used a minimum of 17 sample points for $\rho$ to ensure we faithfully captured the dependence on center-to-center distance, which we expect to play a significant role in determining the forces between particles. We also used a minimum of 2 sample points for two-dimensional nanoparticles and 3 sample points for three-dimensional nanoparticles for the angle coordinates as well as a maximum number of sample points for all coordinates to reduce the size of the search space by excluding interpolants that were unlikely to perform well. All univariate approximations were also required to use nested sample points. Table S1 summarizes the minimum and maximum sample points considered for each coordinate for the different nanoparticles. For the two-dimensional nanoparticles, the approximated pair potential was a function of only three coordinates, $\hat u(\rho, \theta, \alpha)$, so we set the threshold number of sample points to a modest $2 \times 10^3$. For the three-dimensional nanoparticles, we increased $N$ to $10^4$ for the rod, for which $\hat u(\rho, \phi, \alpha, \beta)$ is a function of four coordinates, and to $5 \times 10^4$ for the cube and tetrahedron, for which $\hat u(\rho, \theta, \phi, \alpha, \beta, \gamma)$ depends on all six coordinates.

\begin{figure*}
    \includegraphics{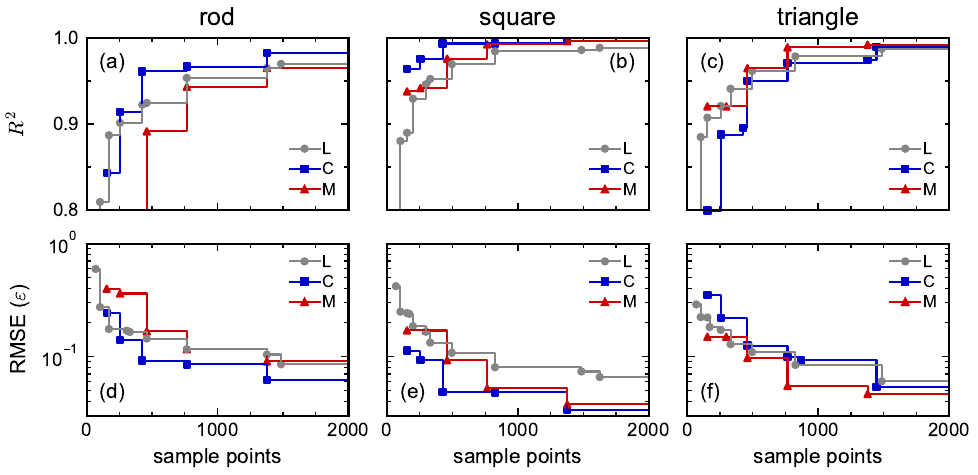}
    \caption{(a--c) $R^2$ and (d--e) RMSE for linear piecewise (L), Chebyshev polynomial (C), and mixed-basis polynomial (M) interpolants for the two-dimensional nanoparticles [(a) \& (d) rod, (b) \& (e) square, and (c) \& (f) triangle] as a function of total number of sample points. For clarity, points are only shown for interpolants that increased $R^2$ or decreased RMSE.}
    \label{fig:error_2d}
\end{figure*}

\begin{figure*}
    \includegraphics{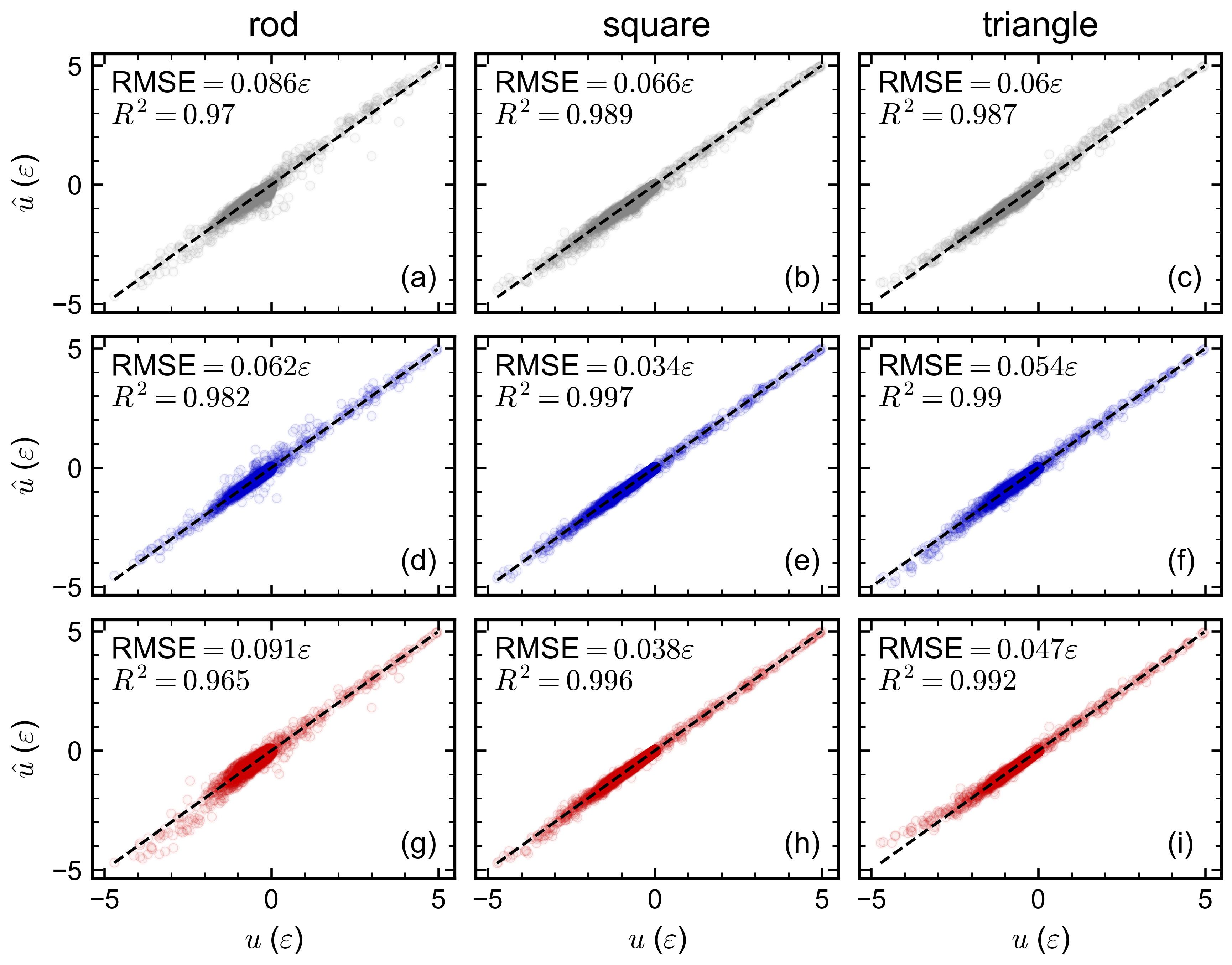}
    \caption{Parity plots of the approximated energy $\hat u$ vs.~the true energy $u$ for the two-dimensional nanoparticles (rod, square, and triangle) and best (a--c) linear piecewise, (d--f) Chebyshev polynomial, and (g--i) mixed-basis polynomial interpolant for each nanoparticle.}
    \label{fig:parity_2d}
\end{figure*}

\begin{figure*}
    \includegraphics{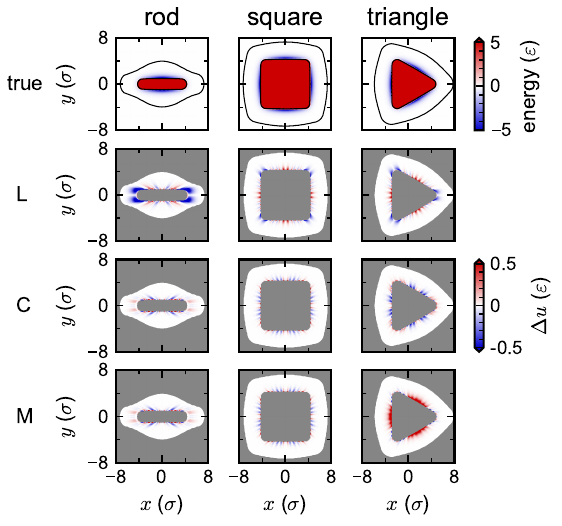}
    \caption{True energy $u$ (top row) and energy residual $\Delta u = \hat u - u$ for the two-dimensional nanoparticles (rod, square, and triangle) and best linear piecewise (L), Chebyshev polynomial (C), and mixed-basis polynomial interpolants. In the top row, the black lines demarcate the domain of center-to-center distances considered for $\hat u$, while in the remaining rows, the grey regions indicate configurations that are outside the domain of $\hat u$.}
    \label{fig:heatmap_2d}
\end{figure*}

To compare the accuracy of the three interpolation schemes, we randomly generated an additional $10^4$ test configurations for the two-dimensional nanoparticles and $5 \times 10^4$ test configurations for the three-dimensional nanoparticles using uniform sampling in the spherical coordinates and Euler angles with rejection to obtain only points that, when reduced, were inside the domain of the interpolants. We computed the square of the Pearson correlation coefficient $R^2$ and root mean square error (RMSE) for each multivariate interpolant by comparing the approximated energy $\hat u$ and the true energy $u$ for the test configurations. For clarity, we will present only the results from interpolants that improved (had larger $R^2$ or smaller RMSE) as the total number of sample points increased. Both metrics identified the same best interpolants, with the exception of the linear piecewise interpolant for the tetrahedron. For most of the results we present here, we computed $r_0$ numerically for both the sample points and test points so that we could focus on the accuracy of $\hat u$. We will discuss approximation of $r_0$ itself at the end of this section.

\subsection{Two-dimensional nanoparticles}

We first calculated $R^2$ and the RMSE for each interpolation scheme for our two-dimensional nanoparticles (Figure \ref{fig:error_2d}). For the rod, the Chebsyhev polynomial interpolants were slightly more accurate than the linear piecewise interpolants, but the mixed-basis polynomial interpolants were slightly less accurate; the Chebyshev polynomial interpolant additionally improved fastest with number of sample points. For the square, both the Chebyshev and mixed-basis polynomial interpolants approximated the energy more accurately and improved faster with number of sample points than the linear piecewise interpolants. For the triangle, all three interpolation schemes reached a similar level of final accuracy, but the mixed-basis polynomial interpolants had the smallest RMSE and improved fastest with number of sample points. The number of univariate sample points used for each coordinate is shown in Table S2 for the best interpolant for each particle type and scheme. We noted that the piecewise linear interpolants consistently required more sample points with respect to $\rho$ than the polynomial interpolants, and as a result, were restricted to use fewer points with respect to the angles $\theta$ and $\alpha$.

To better understand differences in interpolation schemes, we created parity plots of the approximated energy $\hat u$ vs.~the true energy $u$ for the best interpolants found for each scheme (Figure \ref{fig:parity_2d}). In all cases, the parity plots showed quite satisfactory agreement, consistent with $R^2$ being close to one and RMSE being close to zero. The linear piecewise interpolants consistently overpredicted the energy of repulsive configurations somewhat for all particles, while the Chebyshev polynomial and mixed-basis polynomial interpolants were highly accurate for those configurations. This error is likely due to the difference in functional form of the interpolants. All interpolation schemes typically captured the energy of attractive configurations well, but the mixed-basis polynomial interpolant underpredicted the attraction between triangles. These differences can be more clearly seen in a plot of the residual $\Delta u = \hat u - u$ against the true energy $u$ (Figure S2), where systematic deviations of $\Delta u$ from zero suggest there are some behaviors not captured by the approximations.

\begin{figure*}
    \includegraphics{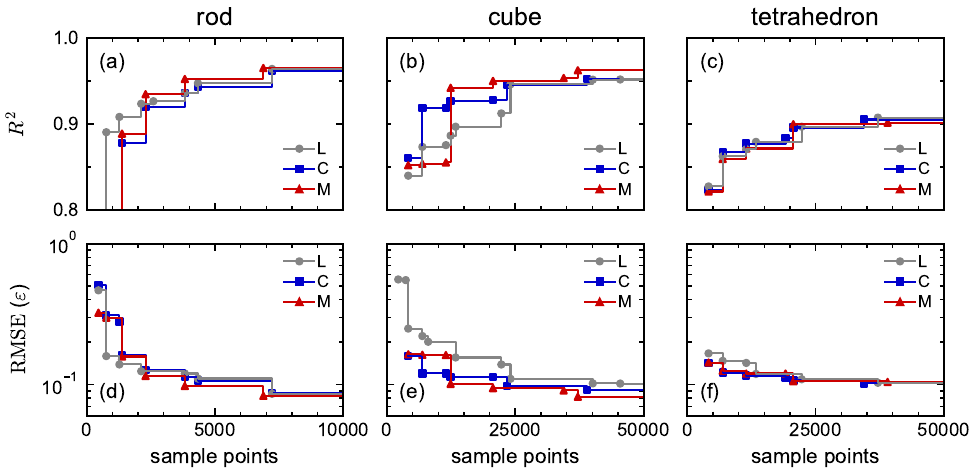}
    \caption{(a--c) $R^2$ and (d--e) RMSE for linear piecewise (L), Chebyshev polynomial (C), and mixed-basis polynomial (M) interpolants for the three-dimensional nanoparticles [(a) \& (d) rod, (b) \& (e) square, and (c) \& (f) triangle] as a function of total number of sample points. For clarity, points are only shown for interpolants that increased $R^2$ or decreased RMSE.}
    \label{fig:error_3d}
\end{figure*}

To more specifically understand the configurations these errors were associated with, we evaluated the energy as a function of relative translation for particles fixed in an orientation that included the minimum-energy configuration. Figure \ref{fig:heatmap_2d} shows the true energy $u$ in the $xy$ plane, along with the lower bound $r_0$ and upper bound $r_0 + r_{\rm c}$ on the center-to-center distance $r$. This visualization confirms that the cutoff we selected was sufficiently large that the true energy was negligible beyond this upper bound. We then evaluated the residual $\Delta u$ for each best interpolant for only configurations inside the domain of $\hat u$. The linear piecewise interpolant struggled for configurations near corners of the particles, with the residual typically being negative for these configurations (i.e., the approximated energy is more attractive than the true energy). The positive residuals for the linear piecewise interpolants tended to occur near close edge-to-edge configurations of the rods, and edge-to-edge configurations of the squares and triangles that tended to be very repulsive. The Chebyshev and mixed-basis polynomial interpolants had mostly similar behavior, showing small inaccuracies for configurations that were close to the minimum-energy configuration. Consistent with the parity plot, the mixed-basis polynomial interpolant, in particular, had a significant positive residual near these attractive configurations. We suspect that the primary reason for this error is that there was no sample point close to this orientation for the triangle ($\alpha = \pi/3$). Traces of the energy as a function of center-to-center distance $r$ for multiple orientations show essentially the same trends (Figure S3).

\subsection{Three-dimensional nanoparticles}
We performed the same analysis of $R^2$ and RMSE for our three-dimensional nanoparticles (Figure \ref{fig:error_3d}). For the rod, all interpolation schemes produced accurate approximations, with the mixed-basis polynomial interpolant having a slightly faster improvement with number of sample points than the other two. We note that substantially more sample points were required to achieve a comparable accuracy as for the two-dimensional rod, which was expected because of the increase in both number of coordinates and total solid angle being approximated for the three-dimensional rod. For the cube, the mixed-basis polynomial interpolant outperformed the linear piecewise and Chebyshev polynomial interpolants, achieving comparable accuracy as for the three-dimensional rod. However, even more sample points were now required. Although the total solid angle was actually smaller for the cube than for the three-dimensional rod (Table \ref{tab:angles}), the cube was described by six coordinates rather than four. Because we used a tensor product to construct our multivariate interpolant, the increase in number of coordinates restricted us to smaller degrees of univariate approximation per coordinate in order to stay below the threshold total number of sample points. This tradeoff highlights one challenge of constructing multivariate interpolants using tensor products, whose size grows highly unfavorably with dimensionality. For the tetrahedron, all interpolation schemes had similar performance but achieved a smaller $R^2$ and larger RMSE than for any of the other nanoparticles. We expected that the tetrahedron would be the most challenging particle to approximate due to its lesser degree of symmetry, and our results suggest that more sample points may be needed to better approximate this interaction.

\begin{figure*}
    \includegraphics{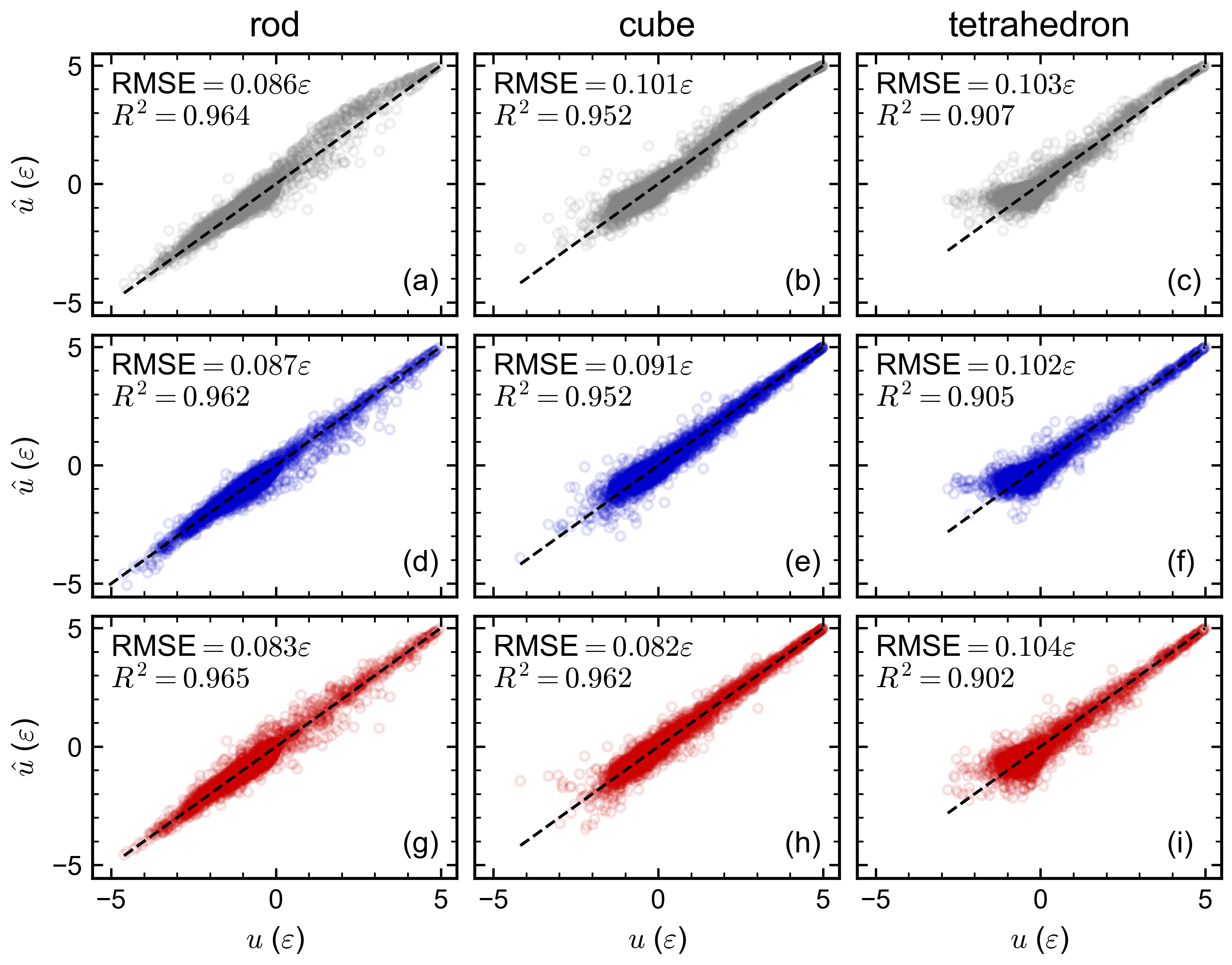}
    \caption{Parity plots of the approximated energy $\hat u$ vs.~the true energy $u$ for the three-dimensional nanoparticles (rod, cube, and tetrahedron) and best (a--c) linear piecewise, (d--f) Chebyshev polynomial, and (g--i) mixed-basis polynomial interpolant for each nanoparticle.}
    \label{fig:parity_3d}
\end{figure*}

As for the two-dimensional nanoparticles, we also created parity plots for the best-performing interpolant for each particle type and interpolation scheme (Figure \ref{fig:parity_3d}). The number of univariate sample points used for each coordinate is shown in Table S2. Qualitatively similar behavior was obtained as for the two-dimensional nanoparticles but the scatter in the points increased, consistent with the somewhat smaller $R^2$ and larger RMSE. One notable difference is that all methods struggled to capture the attractions between tetrahedra, even though the most attractive configurations were not present in the test set. All the interpolation schemes gave systematically positive residuals for these configurations (Figure S4). This systematic error also suggests that these configurations were not adequately sampled by any of the interpolation schemes. Increasing the degree of the univariate approximations would likely help address this issue at the expense of adding additional sample points. Traces of the energy as a function of center-to-center distance $r$ for multiple orientations show essentially the same trends (Figure S5).

\subsection{Approximation of $r_0$}
\begin{figure*}
    \includegraphics{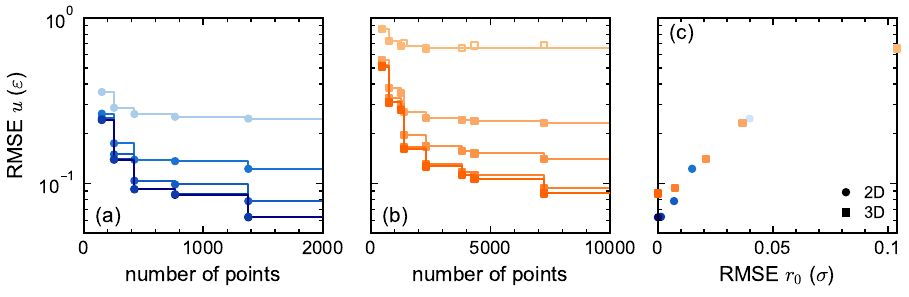}
    \caption{RMSE for approximated energy $\hat{u}$ using approximated $\hat{r}_0$ to evaluate $\rho$ for (a) two-dimensional rod and (b) three-dimensional rod. The different $\hat u$ in (a) and (b) are the same as in Figures \ref{fig:error_2d}(d) and \ref{fig:error_3d}(d), respectively. The number of sample points increases and the RMSE decreases for $\hat r_0$ as the color shade darkens (Table S3). The lightest curve uses the angular points of the best $\hat{u}$ to construct $\hat r_0$, while the darkest curve uses a numerical evaluation of $r_0$. (c) Correlation between RMSE for $\hat u$ and for $\hat r_0$ for the same particles.}
    \label{fig:r0_error}
\end{figure*}
Until now, we have numerically computed $r_0$ each time it was needed to determine $\rho$ so that we could focus on comparing different schemes for approximating $\hat u$. However, for practical purposes in simulations, an approximation of $r_0(\theta, \phi, \boldsymbol{\Omega})$ is also needed for faster evaluation. The multivariate interpolation methods and framework we have used for $\hat u$ are amenable to approximating $r_0$; however, it is not immediately clear how much error in an approximation $\hat r_0$ may affect the accuracy of $\hat u$. We explored this question using the two-dimensional and three-dimensional rod as a case study.

We considered only Chebyshev polynomial interpolants given their good performance for both rods. We first created an approximation $\hat r_0$ using the same angle points as the best Chebyshev polynomial interpolant for each rod, which had a modest RMSE of $0.039\,\sigma$ for the two-dimensional rods and $0.104\,\sigma$ for the three-dimensional rods. We then computed the RMSE for all the Chebyshev polynomial interpolants appearing in Figures \ref{fig:error_2d}(d) and \ref{fig:error_3d}(d), but we now used $\hat r_0$ to calculate $\rho$. The RMSE for $\hat u$ unfortunately increased significantly compared to using the numerically determined $r_0$ (Figure \ref{fig:r0_error}), indicating that having an accurate approximation $\hat r_0$ is likely important. To assess how much $\hat r_0$ needed to improve, we constructed additional approximations using more points for each cooordinate (Table S3), aiming to reduce the RMSE for both $\hat r_0$ and $\hat u$. Indeed, we found that we could achieve comparable accuracy to the numerically determined $r_0$ using a multivariate interpolant $\hat r_0$ with more sample points. We correlated the RMSE for $\hat u$ with the RMSE for $\hat r_0$ [Figure \ref{fig:r0_error}(c)] and found that an RMSE for $\hat r_0$ less than $0.01\,\sigma$ produced comparable RMSE in $\hat u$ as when using the numerically determined $r_0$. This test supports the viability of approximating $\hat r_0$ in addition to $\hat u$.

\section{Conclusions}
In this work, we developed a framework for using multivariate interpolation to approximate anisotropic pair potentials. We focused on multivariate interpolants that used basis functions and sample points constructed from a tensor product of those for univariate Chebyshev and trigonometric polynomials. The key elements of the framework are: (1) a physically-motivated coordinate transformation that excludes highly energetically unfavorable configurations, (2) a systematic reduction of the domain to be approximated using symmetry, (3) refinement of univariate approximations for each coordinate, and (4) use of trigonometric polynomials to enforce differentiability with respect to periodic coordinates. Our tests conducted while developing our framework indicate that the first three elements are particularly important for obtaining good approximations. 

We showed that our approach could accurately approximate the energy between three different two-dimensional nanoparticles using fewer than $2 \times 10^3$ sample points, a three-dimensional rod using fewer than $10^4$ sample points, and a cube using fewer than $5 \times 10^4$ sample points. We also obtained a reasonable approximation of the energy between two tetrahedra; however, it was less accurate than for the other nanoparticles we tested, which we attribute to the tetrahedron having the least symmetry. Overall, our multivariate polynomial interpolants typically performed at least as well as, if not better than, linear piecewise interpolants that we also tested as a reference point (particularly for repulsive configurations) and have the benefit of being differentiable everywhere in the approximation domain. They also required substantially fewer sample points than have been previously used to construct data-driven approximations of anisotropic pair potentials \cite{argun:jchemphys:2024, wilson:jchemphys:2023}.

There is a practical tradeoff between accuracy and  computational cost that must be considered when designing the interpolation scheme. The interpolant is the sum of products of up to six univariate polynomials. As the degree of each univariate approximation and/or the number of coordinates included increases, the number of sample points that must be collected as well as the number of terms that must be evaluated both increase. The samples are only collected once and are all known in advance, so these calculations are trivially parallel and potentially well-suited to high-throughput computing. However, a large number of sample points may be prohibitive to collect if each individual calculation is demanding. Additionally, the increase in the number of terms increases the computational cost of evaluating the interpolant, which may become limiting if done repeatedly in a simulation.  Sparse interpolation \cite{smolyak:dokl:1963, nance:jscicomp:2015, nance:jchemtheroycomp:2014, morrow:jchemtheorycomp:2021, morrow:jscicomp:2020, morrow:jphyschemb:2019, judd:jecondyncon:2014, petix:jchemtheorycomp:2024, kieslich:jgo:2018} is one strategy that can potentially help address this tradeoff by reducing the number of sample points and terms in the multivariate interpolant, which we intend to explore in the future.

\begin{acknowledgments}
We acknowledge support from the Auburn University Research Support Program (M.F.), the National Institutes of Health under Award No.~R35GM147164 (C.A.K.), and the National Science Foundation under Award No.~2223084 (M.P.H). This work was completed with resources provided by the Auburn University Easley Cluster.
\end{acknowledgments}

\bibliography{references}

%aipnum4-2.bst 2019-01-14 (MD) hand-edited version of apsrev4-1.bst
\begin{thebibliography}{49}%
\makeatletter
\providecommand \@ifxundefined [1]{%
 \@ifx{#1\undefined}
}%
\providecommand \@ifnum [1]{%
 \ifnum #1\expandafter \@firstoftwo
 \else \expandafter \@secondoftwo
 \fi
}%
\providecommand \@ifx [1]{%
 \ifx #1\expandafter \@firstoftwo
 \else \expandafter \@secondoftwo
 \fi
}%
\providecommand \natexlab [1]{#1}%
\providecommand \enquote  [1]{``#1''}%
\providecommand \bibnamefont  [1]{#1}%
\providecommand \bibfnamefont [1]{#1}%
\providecommand \citenamefont [1]{#1}%
\providecommand \href@noop [0]{\@secondoftwo}%
\providecommand \href [0]{\begingroup \@sanitize@url \@href}%
\providecommand \@href[1]{\@@startlink{#1}\@@href}%
\providecommand \@@href[1]{\endgroup#1\@@endlink}%
\providecommand \@sanitize@url [0]{\catcode `\\12\catcode `\$12\catcode
  `\&12\catcode `\#12\catcode `\^12\catcode `\_12\catcode `\%12\relax}%
\providecommand \@@startlink[1]{}%
\providecommand \@@endlink[0]{}%
\providecommand \url  [0]{\begingroup\@sanitize@url \@url }%
\providecommand \@url [1]{\endgroup\@href {#1}{\urlprefix }}%
\providecommand \urlprefix  [0]{URL }%
\providecommand \Eprint [0]{\href }%
\providecommand \doibase [0]{https://doi.org/}%
\providecommand \selectlanguage [0]{\@gobble}%
\providecommand \bibinfo  [0]{\@secondoftwo}%
\providecommand \bibfield  [0]{\@secondoftwo}%
\providecommand \translation [1]{[#1]}%
\providecommand \BibitemOpen [0]{}%
\providecommand \bibitemStop [0]{}%
\providecommand \bibitemNoStop [0]{.\EOS\space}%
\providecommand \EOS [0]{\spacefactor3000\relax}%
\providecommand \BibitemShut  [1]{\csname bibitem#1\endcsname}%
\let\auto@bib@innerbib\@empty
%</preamble>
\bibitem [{\citenamefont {Boles}, \citenamefont {Engel},\ and\ \citenamefont
  {Talapin}(2016)}]{boles:chemrev:2016}%
  \BibitemOpen
  \bibfield  {author} {\bibinfo {author} {\bibfnamefont {M.~A.}\ \bibnamefont
  {Boles}}, \bibinfo {author} {\bibfnamefont {M.}~\bibnamefont {Engel}},\ and\
  \bibinfo {author} {\bibfnamefont {D.~V.}\ \bibnamefont {Talapin}},\
  }\bibfield  {title} {\enquote {\bibinfo {title} {Self-assembly of colloidal
  nanocrystals: From intricate structures to functional materials},}\ }\href
  {https://doi.org/10.1021/acs.chemrev.6b00196} {\bibfield  {journal} {\bibinfo
   {journal} {Chemical Reviews}\ }\textbf {\bibinfo {volume} {116}},\ \bibinfo
  {pages} {11220–11289} (\bibinfo {year} {2016})}\BibitemShut {NoStop}%
\bibitem [{\citenamefont {Cai}\ \emph {et~al.}(2021)\citenamefont {Cai},
  \citenamefont {Teng}, \citenamefont {Li}, \citenamefont {Ravaine},
  \citenamefont {He}, \citenamefont {Song}, \citenamefont {Yin}, \citenamefont
  {Zheng},\ and\ \citenamefont {Zhang}}]{cai:chemsocrev:2021}%
  \BibitemOpen
  \bibfield  {author} {\bibinfo {author} {\bibfnamefont {Z.}~\bibnamefont
  {Cai}}, \bibinfo {author} {\bibfnamefont {J.}~\bibnamefont {Teng}}, \bibinfo
  {author} {\bibfnamefont {Z.}~\bibnamefont {Li}}, \bibinfo {author}
  {\bibfnamefont {S.}~\bibnamefont {Ravaine}}, \bibinfo {author} {\bibfnamefont
  {M.}~\bibnamefont {He}}, \bibinfo {author} {\bibfnamefont {Y.}~\bibnamefont
  {Song}}, \bibinfo {author} {\bibfnamefont {Y.}~\bibnamefont {Yin}}, \bibinfo
  {author} {\bibfnamefont {H.}~\bibnamefont {Zheng}},\ and\ \bibinfo {author}
  {\bibfnamefont {A.}~\bibnamefont {Zhang}},\ }\bibfield  {title} {\enquote
  {\bibinfo {title} {From colloidal particles to photonic crystals: Advances in
  self-assembly and their emerging applications},}\ }\href
  {https://doi.org/10.1039/d0cs00706d} {\bibfield  {journal} {\bibinfo
  {journal} {Chemical Society Reviews}\ }\textbf {\bibinfo {volume} {50}},\
  \bibinfo {pages} {5898–5951} (\bibinfo {year} {2021})}\BibitemShut
  {NoStop}%
\bibitem [{\citenamefont {Roth}, \citenamefont {Neal},\ and\ \citenamefont
  {Lenhoff}(1996)}]{roth:biophysj:1996}%
  \BibitemOpen
  \bibfield  {author} {\bibinfo {author} {\bibfnamefont {C.~M.}\ \bibnamefont
  {Roth}}, \bibinfo {author} {\bibfnamefont {B.~L.}\ \bibnamefont {Neal}},\
  and\ \bibinfo {author} {\bibfnamefont {A.~M.}\ \bibnamefont {Lenhoff}},\
  }\bibfield  {title} {\enquote {\bibinfo {title} {Van der waals interactions
  involving proteins},}\ }\href {https://doi.org/10.1016/S0006-3495(96)79641-8}
  {\bibfield  {journal} {\bibinfo  {journal} {Biophysical Journal}\ }\textbf
  {\bibinfo {volume} {70}},\ \bibinfo {pages} {977–987} (\bibinfo {year}
  {1996})}\BibitemShut {NoStop}%
\bibitem [{\citenamefont {Hagan}(2014)}]{hagan:advchemphys:2014}%
  \BibitemOpen
  \bibfield  {author} {\bibinfo {author} {\bibfnamefont {M.~F.}\ \bibnamefont
  {Hagan}},\ }\bibfield  {title} {\enquote {\bibinfo {title} {Modeling viral
  capsid assembly},}\ }in\ \href {https://doi.org/10.1002/9781118755815.ch01}
  {\emph {\bibinfo {booktitle} {Advances in Chemical Physics}}},\ Vol.\
  \bibinfo {volume} {155},\ \bibinfo {editor} {edited by\ \bibinfo {editor}
  {\bibfnamefont {S.~A.}\ \bibnamefont {Rice}}\ and\ \bibinfo {editor}
  {\bibfnamefont {A.~R.}\ \bibnamefont {Dinner}}}\ (\bibinfo  {publisher} {John
  Wiley \& Sons, Inc.},\ \bibinfo {address} {Hoboken, NJ},\ \bibinfo {year}
  {2014})\ pp.\ \bibinfo {pages} {1--67}\BibitemShut {NoStop}%
\bibitem [{\citenamefont {Aumiller}, \citenamefont {Uchida},\ and\
  \citenamefont {Douglas}(2018)}]{aumiller:chemsocrev:2018}%
  \BibitemOpen
  \bibfield  {author} {\bibinfo {author} {\bibfnamefont {W.~M.}\ \bibnamefont
  {Aumiller}}, \bibinfo {author} {\bibfnamefont {M.}~\bibnamefont {Uchida}},\
  and\ \bibinfo {author} {\bibfnamefont {T.}~\bibnamefont {Douglas}},\
  }\bibfield  {title} {\enquote {\bibinfo {title} {Protein cage assembly across
  multiple length scales},}\ }\href {https://doi.org/10.1039/C7CS00818J}
  {\bibfield  {journal} {\bibinfo  {journal} {Chemical Society Reviews}\
  }\textbf {\bibinfo {volume} {47}},\ \bibinfo {pages} {3433--3469} (\bibinfo
  {year} {2018})}\BibitemShut {NoStop}%
\bibitem [{\citenamefont {Solomon}(2011)}]{solomon:colloidandinterface:2011}%
  \BibitemOpen
  \bibfield  {author} {\bibinfo {author} {\bibfnamefont {M.~J.}\ \bibnamefont
  {Solomon}},\ }\bibfield  {title} {\enquote {\bibinfo {title} {Directions for
  targeted self-assembly of anisotropic colloids from statistical
  thermodynamics},}\ }\href
  {https://doi.org/https://doi.org/10.1016/j.cocis.2011.01.006} {\bibfield
  {journal} {\bibinfo  {journal} {Current Opinion in Colloid \& Interface
  Science}\ }\textbf {\bibinfo {volume} {16}},\ \bibinfo {pages} {158--167}
  (\bibinfo {year} {2011})}\BibitemShut {NoStop}%
\bibitem [{\citenamefont {Fejer}, \citenamefont {Chakrabarti},\ and\
  \citenamefont {Wales}(2011)}]{fejer:softmatter:2011}%
  \BibitemOpen
  \bibfield  {author} {\bibinfo {author} {\bibfnamefont {S.~N.}\ \bibnamefont
  {Fejer}}, \bibinfo {author} {\bibfnamefont {D.}~\bibnamefont {Chakrabarti}},\
  and\ \bibinfo {author} {\bibfnamefont {D.~J.}\ \bibnamefont {Wales}},\
  }\bibfield  {title} {\enquote {\bibinfo {title} {Self-assembly of anisotropic
  particles},}\ }\href {https://doi.org/10.1039/C0SM01289K} {\bibfield
  {journal} {\bibinfo  {journal} {Soft Matter}\ }\textbf {\bibinfo {volume}
  {7}},\ \bibinfo {pages} {3553--3564} (\bibinfo {year} {2011})}\BibitemShut
  {NoStop}%
\bibitem [{\citenamefont {Thorkelsson}, \citenamefont {Bai},\ and\
  \citenamefont {Xu}(2015)}]{thorkelsson:nanotoday:2015}%
  \BibitemOpen
  \bibfield  {author} {\bibinfo {author} {\bibfnamefont {K.}~\bibnamefont
  {Thorkelsson}}, \bibinfo {author} {\bibfnamefont {P.}~\bibnamefont {Bai}},\
  and\ \bibinfo {author} {\bibfnamefont {T.}~\bibnamefont {Xu}},\ }\bibfield
  {title} {\enquote {\bibinfo {title} {Self-assembly and applications of
  anisotropic nanomaterials: A review},}\ }\href
  {https://doi.org/https://doi.org/10.1016/j.nantod.2014.12.005} {\bibfield
  {journal} {\bibinfo  {journal} {Nano Today}\ }\textbf {\bibinfo {volume}
  {10}},\ \bibinfo {pages} {48--66} (\bibinfo {year} {2015})}\BibitemShut
  {NoStop}%
\bibitem [{\citenamefont {Meseguer}(2005)}]{meseguer:colsurf:2005}%
  \BibitemOpen
  \bibfield  {author} {\bibinfo {author} {\bibfnamefont {F.}~\bibnamefont
  {Meseguer}},\ }\bibfield  {title} {\enquote {\bibinfo {title} {Colloidal
  crystals as photonic crystals},}\ }\href
  {https://doi.org/10.1016/j.colsurfa.2005.05.038} {\bibfield  {journal}
  {\bibinfo  {journal} {Colloids and Surfaces A: Physicochemical and
  Engineering Aspects}\ }\textbf {\bibinfo {volume} {270–271}},\ \bibinfo
  {pages} {1–7} (\bibinfo {year} {2005})}\BibitemShut {NoStop}%
\bibitem [{\citenamefont {Kim}\ \emph {et~al.}(2011)\citenamefont {Kim},
  \citenamefont {Lee}, \citenamefont {Yang},\ and\ \citenamefont
  {Yi}}]{kim:npasia:2011}%
  \BibitemOpen
  \bibfield  {author} {\bibinfo {author} {\bibfnamefont {S.-H.}\ \bibnamefont
  {Kim}}, \bibinfo {author} {\bibfnamefont {S.~Y.}\ \bibnamefont {Lee}},
  \bibinfo {author} {\bibfnamefont {S.-M.}\ \bibnamefont {Yang}},\ and\
  \bibinfo {author} {\bibfnamefont {G.-R.}\ \bibnamefont {Yi}},\ }\bibfield
  {title} {\enquote {\bibinfo {title} {Self-assembled colloidal structures for
  photonics},}\ }\href {https://doi.org/10.1038/asiamat.2010.192} {\bibfield
  {journal} {\bibinfo  {journal} {NPG Asia Materials}\ }\textbf {\bibinfo
  {volume} {3}},\ \bibinfo {pages} {25–33} (\bibinfo {year}
  {2011})}\BibitemShut {NoStop}%
\bibitem [{\citenamefont {Henzie}\ \emph {et~al.}(2012)\citenamefont {Henzie},
  \citenamefont {Grünwald}, \citenamefont {Widmer-Cooper}, \citenamefont
  {Geissler},\ and\ \citenamefont {Yang}}]{henzie:natmat:2012}%
  \BibitemOpen
  \bibfield  {author} {\bibinfo {author} {\bibfnamefont {J.}~\bibnamefont
  {Henzie}}, \bibinfo {author} {\bibfnamefont {M.}~\bibnamefont {Grünwald}},
  \bibinfo {author} {\bibfnamefont {A.}~\bibnamefont {Widmer-Cooper}}, \bibinfo
  {author} {\bibfnamefont {P.~L.}\ \bibnamefont {Geissler}},\ and\ \bibinfo
  {author} {\bibfnamefont {P.}~\bibnamefont {Yang}},\ }\bibfield  {title}
  {\enquote {\bibinfo {title} {Self-assembly of uniform polyhedral silver
  nanocrystals into densest packings and exotic superlattices},}\ }\href
  {https://doi.org/10.1038/nmat3178} {\bibfield  {journal} {\bibinfo  {journal}
  {Nature Materials}\ }\textbf {\bibinfo {volume} {11}},\ \bibinfo {pages}
  {131–137} (\bibinfo {year} {2012})}\BibitemShut {NoStop}%
\bibitem [{\citenamefont {Yetkin}\ \emph {et~al.}(2024)\citenamefont {Yetkin},
  \citenamefont {Wani}, \citenamefont {Kritika}, \citenamefont {Howard},
  \citenamefont {Kappl}, \citenamefont {Butt},\ and\ \citenamefont
  {Nikoubashman}}]{yektin:langmuir:2024}%
  \BibitemOpen
  \bibfield  {author} {\bibinfo {author} {\bibfnamefont {M.}~\bibnamefont
  {Yetkin}}, \bibinfo {author} {\bibfnamefont {Y.~M.}\ \bibnamefont {Wani}},
  \bibinfo {author} {\bibfnamefont {K.}~\bibnamefont {Kritika}}, \bibinfo
  {author} {\bibfnamefont {M.~P.}\ \bibnamefont {Howard}}, \bibinfo {author}
  {\bibfnamefont {M.}~\bibnamefont {Kappl}}, \bibinfo {author} {\bibfnamefont
  {H.-J.}\ \bibnamefont {Butt}},\ and\ \bibinfo {author} {\bibfnamefont
  {A.}~\bibnamefont {Nikoubashman}},\ }\bibfield  {title} {\enquote {\bibinfo
  {title} {Structure formation in supraparticles composed of spherical and
  elongated particles},}\ }\href {https://doi.org/10.1021/acs.langmuir.3c03410}
  {\bibfield  {journal} {\bibinfo  {journal} {Langmuir}\ }\textbf {\bibinfo
  {volume} {40}},\ \bibinfo {pages} {1096--1108} (\bibinfo {year}
  {2024})}\BibitemShut {NoStop}%
\bibitem [{\citenamefont {Chen}\ \emph {et~al.}(2023)\citenamefont {Chen},
  \citenamefont {Zhang}, \citenamefont {Liu}, \citenamefont {Zang},
  \citenamefont {Lv}, \citenamefont {Du},\ and\ \citenamefont
  {Zhao}}]{chen:advsci:2023}%
  \BibitemOpen
  \bibfield  {author} {\bibinfo {author} {\bibfnamefont {X.}~\bibnamefont
  {Chen}}, \bibinfo {author} {\bibfnamefont {T.}~\bibnamefont {Zhang}},
  \bibinfo {author} {\bibfnamefont {H.}~\bibnamefont {Liu}}, \bibinfo {author}
  {\bibfnamefont {J.}~\bibnamefont {Zang}}, \bibinfo {author} {\bibfnamefont
  {C.}~\bibnamefont {Lv}}, \bibinfo {author} {\bibfnamefont {M.}~\bibnamefont
  {Du}},\ and\ \bibinfo {author} {\bibfnamefont {G.}~\bibnamefont {Zhao}},\
  }\bibfield  {title} {\enquote {\bibinfo {title} {Shape-anisotropic assembly
  of protein nanocages with identical building blocks by designed
  intermolecular $\pi$–$\pi$ interactions},}\ }\href
  {https://doi.org/10.1002/advs.202305398} {\bibfield  {journal} {\bibinfo
  {journal} {Advanced Science}\ }\textbf {\bibinfo {volume} {10}},\ \bibinfo
  {pages} {2305398} (\bibinfo {year} {2023})}\BibitemShut {NoStop}%
\bibitem [{\citenamefont {Wang}\ \emph {et~al.}(2021)\citenamefont {Wang},
  \citenamefont {Bolan}, \citenamefont {Tsang}, \citenamefont {Sarkar},
  \citenamefont {Bradney},\ and\ \citenamefont {Li}}]{wang:jhazmat:2021}%
  \BibitemOpen
  \bibfield  {author} {\bibinfo {author} {\bibfnamefont {X.}~\bibnamefont
  {Wang}}, \bibinfo {author} {\bibfnamefont {N.}~\bibnamefont {Bolan}},
  \bibinfo {author} {\bibfnamefont {D.~C.}\ \bibnamefont {Tsang}}, \bibinfo
  {author} {\bibfnamefont {B.}~\bibnamefont {Sarkar}}, \bibinfo {author}
  {\bibfnamefont {L.}~\bibnamefont {Bradney}},\ and\ \bibinfo {author}
  {\bibfnamefont {Y.}~\bibnamefont {Li}},\ }\bibfield  {title} {\enquote
  {\bibinfo {title} {A review of microplastics aggregation in aquatic
  environment: Influence factors, analytical methods, and environmental
  implications},}\ }\href
  {https://doi.org/https://doi.org/10.1016/j.jhazmat.2020.123496} {\bibfield
  {journal} {\bibinfo  {journal} {Journal of Hazardous Materials}\ }\textbf
  {\bibinfo {volume} {402}},\ \bibinfo {pages} {123496} (\bibinfo {year}
  {2021})}\BibitemShut {NoStop}%
\bibitem [{\citenamefont {Argun}\ and\ \citenamefont
  {Statt}(2023)}]{argun:softmatter:2023}%
  \BibitemOpen
  \bibfield  {author} {\bibinfo {author} {\bibfnamefont {B.~R.}\ \bibnamefont
  {Argun}}\ and\ \bibinfo {author} {\bibfnamefont {A.}~\bibnamefont {Statt}},\
  }\bibfield  {title} {\enquote {\bibinfo {title} {Influence of shape on
  heteroaggregation of model microplastics: a simulation study},}\ }\href
  {https://doi.org/10.1039/D3SM01014G} {\bibfield  {journal} {\bibinfo
  {journal} {Soft Matter}\ }\textbf {\bibinfo {volume} {19}},\ \bibinfo {pages}
  {8081--8090} (\bibinfo {year} {2023})}\BibitemShut {NoStop}%
\bibitem [{\citenamefont {Allen}\ and\ \citenamefont
  {Tildesley}(2017)}]{allen:oxford:2017}%
  \BibitemOpen
  \bibfield  {author} {\bibinfo {author} {\bibfnamefont {M.~P.}\ \bibnamefont
  {Allen}}\ and\ \bibinfo {author} {\bibfnamefont {D.~J.}\ \bibnamefont
  {Tildesley}},\ }\href {https://doi.org/10.1093/oso/9780198803195.001.0001}
  {\emph {\bibinfo {title} {Computer Simulation of Liquids}}}\ (\bibinfo
  {publisher} {Oxford University Press},\ \bibinfo {year} {2017})\BibitemShut
  {NoStop}%
\bibitem [{\citenamefont {Kern}\ and\ \citenamefont
  {Frenkel}(2003)}]{kern:jchemphys:2003}%
  \BibitemOpen
  \bibfield  {author} {\bibinfo {author} {\bibfnamefont {N.}~\bibnamefont
  {Kern}}\ and\ \bibinfo {author} {\bibfnamefont {D.}~\bibnamefont {Frenkel}},\
  }\bibfield  {title} {\enquote {\bibinfo {title} {Fluid–fluid coexistence in
  colloidal systems with short-ranged strongly directional attraction},}\
  }\href {https://doi.org/10.1063/1.1569473} {\bibfield  {journal} {\bibinfo
  {journal} {The Journal of Chemical Physics}\ }\textbf {\bibinfo {volume}
  {118}},\ \bibinfo {pages} {9882--9889} (\bibinfo {year} {2003})}\BibitemShut
  {NoStop}%
\bibitem [{\citenamefont {Berardi}, \citenamefont {Fava},\ and\ \citenamefont
  {Zannoni}(1998)}]{berardi:chemphyslett:1998}%
  \BibitemOpen
  \bibfield  {author} {\bibinfo {author} {\bibfnamefont {R.}~\bibnamefont
  {Berardi}}, \bibinfo {author} {\bibfnamefont {C.}~\bibnamefont {Fava}},\ and\
  \bibinfo {author} {\bibfnamefont {C.}~\bibnamefont {Zannoni}},\ }\bibfield
  {title} {\enquote {\bibinfo {title} {A gay–berne potential for dissimilar
  biaxial particles},}\ }\href {https://doi.org/10.1016/S0009-2614(98)01090-2}
  {\bibfield  {journal} {\bibinfo  {journal} {Chemical Physics Letters}\
  }\textbf {\bibinfo {volume} {297}},\ \bibinfo {pages} {8–14} (\bibinfo
  {year} {1998})}\BibitemShut {NoStop}%
\bibitem [{\citenamefont {Weeks}, \citenamefont {Chandler},\ and\ \citenamefont
  {Andersen}(1971)}]{weeks:jchemphys:1971}%
  \BibitemOpen
  \bibfield  {author} {\bibinfo {author} {\bibfnamefont {J.~D.}\ \bibnamefont
  {Weeks}}, \bibinfo {author} {\bibfnamefont {D.}~\bibnamefont {Chandler}},\
  and\ \bibinfo {author} {\bibfnamefont {H.~C.}\ \bibnamefont {Andersen}},\
  }\bibfield  {title} {\enquote {\bibinfo {title} {Role of repulsive forces in
  determining the equilibrium structure of simple liquids},}\ }\href
  {https://doi.org/10.1063/1.1674820} {\bibfield  {journal} {\bibinfo
  {journal} {The Journal of Chemical Physics}\ }\textbf {\bibinfo {volume}
  {54}},\ \bibinfo {pages} {5237--5247} (\bibinfo {year} {1971})}\BibitemShut
  {NoStop}%
\bibitem [{\citenamefont {Ramasubramani}\ \emph {et~al.}(2020)\citenamefont
  {Ramasubramani}, \citenamefont {Vo}, \citenamefont {Anderson},\ and\
  \citenamefont {Glotzer}}]{ramasubramani:jchemphys:2020}%
  \BibitemOpen
  \bibfield  {author} {\bibinfo {author} {\bibfnamefont {V.}~\bibnamefont
  {Ramasubramani}}, \bibinfo {author} {\bibfnamefont {T.}~\bibnamefont {Vo}},
  \bibinfo {author} {\bibfnamefont {J.~A.}\ \bibnamefont {Anderson}},\ and\
  \bibinfo {author} {\bibfnamefont {S.~C.}\ \bibnamefont {Glotzer}},\
  }\bibfield  {title} {\enquote {\bibinfo {title} {A mean-field approach to
  simulating anisotropic particles},}\ }\href
  {https://doi.org/10.1063/5.0019735} {\bibfield  {journal} {\bibinfo
  {journal} {The Journal of Chemical Physics}\ }\textbf {\bibinfo {volume}
  {153}},\ \bibinfo {pages} {084106} (\bibinfo {year} {2020})}\BibitemShut
  {NoStop}%
\bibitem [{\citenamefont {Nguyen}\ and\ \citenamefont
  {Plimpton}(2019)}]{nguyen:cpc:2019}%
  \BibitemOpen
  \bibfield  {author} {\bibinfo {author} {\bibfnamefont {T.~D.}\ \bibnamefont
  {Nguyen}}\ and\ \bibinfo {author} {\bibfnamefont {S.~J.}\ \bibnamefont
  {Plimpton}},\ }\bibfield  {title} {\enquote {\bibinfo {title} {Aspherical
  particle models for molecular dynamics simulation},}\ }\href
  {https://doi.org/10.1016/j.cpc.2019.05.010} {\bibfield  {journal} {\bibinfo
  {journal} {Computer Physics Communications}\ }\textbf {\bibinfo {volume}
  {243}},\ \bibinfo {pages} {12–24} (\bibinfo {year} {2019})}\BibitemShut
  {NoStop}%
\bibitem [{\citenamefont {Wani}\ \emph {et~al.}(2024)\citenamefont {Wani},
  \citenamefont {Kovakas}, \citenamefont {Nikoubashman},\ and\ \citenamefont
  {Howard}}]{wani:softmatter:2024}%
  \BibitemOpen
  \bibfield  {author} {\bibinfo {author} {\bibfnamefont {Y.~M.}\ \bibnamefont
  {Wani}}, \bibinfo {author} {\bibfnamefont {P.~G.}\ \bibnamefont {Kovakas}},
  \bibinfo {author} {\bibfnamefont {A.}~\bibnamefont {Nikoubashman}},\ and\
  \bibinfo {author} {\bibfnamefont {M.~P.}\ \bibnamefont {Howard}},\ }\bibfield
   {title} {\enquote {\bibinfo {title} {Mesoscale simulations of diffusion and
  sedimentation in shape-anisotropic nanoparticle suspensions},}\ }\href
  {https://doi.org/10.1039/d4sm00271g} {\bibfield  {journal} {\bibinfo
  {journal} {Soft Matter}\ }\textbf {\bibinfo {volume} {20}},\ \bibinfo {pages}
  {3942–3953} (\bibinfo {year} {2024})}\BibitemShut {NoStop}%
\bibitem [{\citenamefont {Nguyen}\ and\ \citenamefont
  {Huang}(2022)}]{nguyen:jchemphys:2022}%
  \BibitemOpen
  \bibfield  {author} {\bibinfo {author} {\bibfnamefont {H.~T.~L.}\
  \bibnamefont {Nguyen}}\ and\ \bibinfo {author} {\bibfnamefont {D.~M.}\
  \bibnamefont {Huang}},\ }\bibfield  {title} {\enquote {\bibinfo {title}
  {Systematic bottom-up molecular coarse-graining via force and torque matching
  using anisotropic particles},}\ }\href {https://doi.org/10.1063/5.0085006}
  {\bibfield  {journal} {\bibinfo  {journal} {The Journal of Chemical Physics}\
  }\textbf {\bibinfo {volume} {156}},\ \bibinfo {pages} {184118} (\bibinfo
  {year} {2022})}\BibitemShut {NoStop}%
\bibitem [{\citenamefont {Noid}\ \emph
  {et~al.}(2008{\natexlab{a}})\citenamefont {Noid}, \citenamefont {Chu},
  \citenamefont {Ayton}, \citenamefont {Krishna}, \citenamefont {Izvekov},
  \citenamefont {Voth}, \citenamefont {Das},\ and\ \citenamefont
  {Andersen}}]{noid:jchemphys:2008_1}%
  \BibitemOpen
  \bibfield  {author} {\bibinfo {author} {\bibfnamefont {W.~G.}\ \bibnamefont
  {Noid}}, \bibinfo {author} {\bibfnamefont {J.-W.}\ \bibnamefont {Chu}},
  \bibinfo {author} {\bibfnamefont {G.~S.}\ \bibnamefont {Ayton}}, \bibinfo
  {author} {\bibfnamefont {V.}~\bibnamefont {Krishna}}, \bibinfo {author}
  {\bibfnamefont {S.}~\bibnamefont {Izvekov}}, \bibinfo {author} {\bibfnamefont
  {G.~A.}\ \bibnamefont {Voth}}, \bibinfo {author} {\bibfnamefont
  {A.}~\bibnamefont {Das}},\ and\ \bibinfo {author} {\bibfnamefont {H.~C.}\
  \bibnamefont {Andersen}},\ }\bibfield  {title} {\enquote {\bibinfo {title}
  {The multiscale coarse-graining method. i. a rigorous bridge between
  atomistic and coarse-grained models},}\ }\href
  {https://doi.org/10.1063/1.2938860} {\bibfield  {journal} {\bibinfo
  {journal} {The Journal of Chemical Physics}\ }\textbf {\bibinfo {volume}
  {128}},\ \bibinfo {pages} {244114} (\bibinfo {year}
  {2008}{\natexlab{a}})}\BibitemShut {NoStop}%
\bibitem [{\citenamefont {Noid}\ \emph
  {et~al.}(2008{\natexlab{b}})\citenamefont {Noid}, \citenamefont {Liu},
  \citenamefont {Wang}, \citenamefont {Chu}, \citenamefont {Ayton},
  \citenamefont {Izvekov}, \citenamefont {Andersen},\ and\ \citenamefont
  {Voth}}]{noid:jchemphys:2008_2}%
  \BibitemOpen
  \bibfield  {author} {\bibinfo {author} {\bibfnamefont {W.~G.}\ \bibnamefont
  {Noid}}, \bibinfo {author} {\bibfnamefont {P.}~\bibnamefont {Liu}}, \bibinfo
  {author} {\bibfnamefont {Y.}~\bibnamefont {Wang}}, \bibinfo {author}
  {\bibfnamefont {J.-W.}\ \bibnamefont {Chu}}, \bibinfo {author} {\bibfnamefont
  {G.~S.}\ \bibnamefont {Ayton}}, \bibinfo {author} {\bibfnamefont
  {S.}~\bibnamefont {Izvekov}}, \bibinfo {author} {\bibfnamefont {H.~C.}\
  \bibnamefont {Andersen}},\ and\ \bibinfo {author} {\bibfnamefont {G.~A.}\
  \bibnamefont {Voth}},\ }\bibfield  {title} {\enquote {\bibinfo {title} {The
  multiscale coarse-graining method. ii. numerical implementation for
  coarse-grained molecular models},}\ }\href
  {https://doi.org/10.1063/1.2938857} {\bibfield  {journal} {\bibinfo
  {journal} {The Journal of Chemical Physics}\ }\textbf {\bibinfo {volume}
  {128}},\ \bibinfo {pages} {244115} (\bibinfo {year}
  {2008}{\natexlab{b}})}\BibitemShut {NoStop}%
\bibitem [{\citenamefont {Campos-Villalobos}\ \emph {et~al.}(2022)\citenamefont
  {Campos-Villalobos}, \citenamefont {Giunta}, \citenamefont {Marín-Aguilar},\
  and\ \citenamefont {Dijkstra}}]{campos-villalobos:jchemphys:2022}%
  \BibitemOpen
  \bibfield  {author} {\bibinfo {author} {\bibfnamefont {G.}~\bibnamefont
  {Campos-Villalobos}}, \bibinfo {author} {\bibfnamefont {G.}~\bibnamefont
  {Giunta}}, \bibinfo {author} {\bibfnamefont {S.}~\bibnamefont
  {Marín-Aguilar}},\ and\ \bibinfo {author} {\bibfnamefont {M.}~\bibnamefont
  {Dijkstra}},\ }\bibfield  {title} {\enquote {\bibinfo {title}
  {Machine-learning effective many-body potentials for anisotropic particles
  using orientation-dependent symmetry functions},}\ }\href
  {https://doi.org/10.1063/5.0091319} {\bibfield  {journal} {\bibinfo
  {journal} {The Journal of Chemical Physics}\ }\textbf {\bibinfo {volume}
  {157}},\ \bibinfo {pages} {024902} (\bibinfo {year} {2022})}\BibitemShut
  {NoStop}%
\bibitem [{\citenamefont {Campos-Villalobos}\ \emph {et~al.}(2024)\citenamefont
  {Campos-Villalobos}, \citenamefont {Subert}, \citenamefont {Giunta},\ and\
  \citenamefont {Dijkstra}}]{campos-villalobos:npjcompmat:2024}%
  \BibitemOpen
  \bibfield  {author} {\bibinfo {author} {\bibfnamefont {G.}~\bibnamefont
  {Campos-Villalobos}}, \bibinfo {author} {\bibfnamefont {R.}~\bibnamefont
  {Subert}}, \bibinfo {author} {\bibfnamefont {G.}~\bibnamefont {Giunta}},\
  and\ \bibinfo {author} {\bibfnamefont {M.}~\bibnamefont {Dijkstra}},\
  }\bibfield  {title} {\enquote {\bibinfo {title} {Machine-learned
  coarse-grained potentials for particles with anisotropic shapes and
  interactions},}\ }\href {https://doi.org/10.1038/s41524-024-01405-4}
  {\bibfield  {journal} {\bibinfo  {journal} {npj Computational Materials}\
  }\textbf {\bibinfo {volume} {10}},\ \bibinfo {pages} {228} (\bibinfo {year}
  {2024})}\BibitemShut {NoStop}%
\bibitem [{\citenamefont {Lin}\ \emph {et~al.}(2024)\citenamefont {Lin},
  \citenamefont {Huguenin-Dumittan}, \citenamefont {Cho}, \citenamefont
  {Nigam},\ and\ \citenamefont {Cersonsky}}]{lin:jchemphys:2024}%
  \BibitemOpen
  \bibfield  {author} {\bibinfo {author} {\bibfnamefont {A.}~\bibnamefont
  {Lin}}, \bibinfo {author} {\bibfnamefont {K.~K.}\ \bibnamefont
  {Huguenin-Dumittan}}, \bibinfo {author} {\bibfnamefont {Y.-C.}\ \bibnamefont
  {Cho}}, \bibinfo {author} {\bibfnamefont {J.}~\bibnamefont {Nigam}},\ and\
  \bibinfo {author} {\bibfnamefont {R.~K.}\ \bibnamefont {Cersonsky}},\
  }\bibfield  {title} {\enquote {\bibinfo {title} {Expanding
  density-correlation machine learning representations for anisotropic
  coarse-grained particles},}\ }\href {https://doi.org/10.1063/5.0210910}
  {\bibfield  {journal} {\bibinfo  {journal} {The Journal of Chemical Physics}\
  }\textbf {\bibinfo {volume} {161}},\ \bibinfo {pages} {074112} (\bibinfo
  {year} {2024})}\BibitemShut {NoStop}%
\bibitem [{\citenamefont {Wilson}\ and\ \citenamefont
  {Huang}(2023)}]{wilson:jchemphys:2023}%
  \BibitemOpen
  \bibfield  {author} {\bibinfo {author} {\bibfnamefont {M.~O.}\ \bibnamefont
  {Wilson}}\ and\ \bibinfo {author} {\bibfnamefont {D.~M.}\ \bibnamefont
  {Huang}},\ }\bibfield  {title} {\enquote {\bibinfo {title} {Anisotropic
  molecular coarse-graining by force and torque matching with neural
  networks},}\ }\href {https://doi.org/10.1063/5.0143724} {\bibfield  {journal}
  {\bibinfo  {journal} {The Journal of Chemical Physics}\ }\textbf {\bibinfo
  {volume} {159}},\ \bibinfo {pages} {024110} (\bibinfo {year}
  {2023})}\BibitemShut {NoStop}%
\bibitem [{\citenamefont {Argun}, \citenamefont {Fu},\ and\ \citenamefont
  {Statt}(2024)}]{argun:jchemphys:2024}%
  \BibitemOpen
  \bibfield  {author} {\bibinfo {author} {\bibfnamefont {B.~R.}\ \bibnamefont
  {Argun}}, \bibinfo {author} {\bibfnamefont {Y.}~\bibnamefont {Fu}},\ and\
  \bibinfo {author} {\bibfnamefont {A.}~\bibnamefont {Statt}},\ }\bibfield
  {title} {\enquote {\bibinfo {title} {Molecular dynamics simulations of
  anisotropic particles accelerated by neural-net predicted interactions},}\
  }\href {https://doi.org/10.1063/5.0206636} {\bibfield  {journal} {\bibinfo
  {journal} {The Journal of Chemical Physics}\ }\textbf {\bibinfo {volume}
  {160}},\ \bibinfo {pages} {244901} (\bibinfo {year} {2024})}\BibitemShut
  {NoStop}%
\bibitem [{\citenamefont {Press}\ \emph {et~al.}(2007)\citenamefont {Press},
  \citenamefont {Teukolsky}, \citenamefont {Vetterling},\ and\ \citenamefont
  {Flannery}}]{press:cambridge:2007}%
  \BibitemOpen
  \bibfield  {author} {\bibinfo {author} {\bibfnamefont {W.~H.}\ \bibnamefont
  {Press}}, \bibinfo {author} {\bibfnamefont {S.~A.}\ \bibnamefont
  {Teukolsky}}, \bibinfo {author} {\bibfnamefont {W.~T.}\ \bibnamefont
  {Vetterling}},\ and\ \bibinfo {author} {\bibfnamefont {B.~P.}\ \bibnamefont
  {Flannery}},\ }\href@noop {} {\emph {\bibinfo {title} {Numerical Recipes 3rd
  Edition: The Art of Scientific Computing}}},\ \bibinfo {edition} {3rd}\ ed.\
  (\bibinfo  {publisher} {Cambridge University Press},\ \bibinfo {address}
  {USA},\ \bibinfo {year} {2007})\ Chap.~\bibinfo {chapter} {5}, pp.\ \bibinfo
  {pages} {201--254}\BibitemShut {NoStop}%
\bibitem [{\citenamefont {Trefethen}(2019)}]{trefethen:socindmath:2019}%
  \BibitemOpen
  \bibfield  {author} {\bibinfo {author} {\bibfnamefont {L.~N.}\ \bibnamefont
  {Trefethen}},\ }\href {https://doi.org/10.1137/1.9781611975949} {\emph
  {\bibinfo {title} {Approximation Theory and Approximation Practice, Extended
  Edition}}}\ (\bibinfo  {publisher} {Society for Industrial and Applied
  Mathematics},\ \bibinfo {address} {Philadelphia, PA},\ \bibinfo {year}
  {2019})\ Chap.~\bibinfo {chapter} {2}, pp.\ \bibinfo {pages} {7--12},\
  \Eprint
  {https://arxiv.org/abs/https://epubs.siam.org/doi/pdf/10.1137/1.9781611975949}
  {https://epubs.siam.org/doi/pdf/10.1137/1.9781611975949} \BibitemShut
  {NoStop}%
\bibitem [{\citenamefont {Lindsey}, \citenamefont {Fried},\ and\ \citenamefont
  {Goldman}(2017)}]{lindsey:jchemtheorycomp:2017}%
  \BibitemOpen
  \bibfield  {author} {\bibinfo {author} {\bibfnamefont {R.~K.}\ \bibnamefont
  {Lindsey}}, \bibinfo {author} {\bibfnamefont {L.~E.}\ \bibnamefont {Fried}},\
  and\ \bibinfo {author} {\bibfnamefont {N.}~\bibnamefont {Goldman}},\
  }\bibfield  {title} {\enquote {\bibinfo {title} {Chimes: A force matched
  potential with explicit three-body interactions for molten carbon},}\ }\href
  {https://doi.org/10.1021/acs.jctc.7b00867} {\bibfield  {journal} {\bibinfo
  {journal} {Journal of Chemical Theory and Computation}\ }\textbf {\bibinfo
  {volume} {13}},\ \bibinfo {pages} {6222--6229} (\bibinfo {year}
  {2017})}\BibitemShut {NoStop}%
\bibitem [{\citenamefont {Lindsey}, \citenamefont {Fried},\ and\ \citenamefont
  {Goldman}(2019)}]{lindsey:jchemtheorycomp:2019}%
  \BibitemOpen
  \bibfield  {author} {\bibinfo {author} {\bibfnamefont {R.~K.}\ \bibnamefont
  {Lindsey}}, \bibinfo {author} {\bibfnamefont {L.~E.}\ \bibnamefont {Fried}},\
  and\ \bibinfo {author} {\bibfnamefont {N.}~\bibnamefont {Goldman}},\
  }\bibfield  {title} {\enquote {\bibinfo {title} {Application of the chimes
  force field to nonreactive molecular systems: Water at ambient conditions},}\
  }\href {https://doi.org/10.1021/acs.jctc.8b00831} {\bibfield  {journal}
  {\bibinfo  {journal} {Journal of Chemical Theory and Computation}\ }\textbf
  {\bibinfo {volume} {15}},\ \bibinfo {pages} {436--447} (\bibinfo {year}
  {2019})}\BibitemShut {NoStop}%
\bibitem [{\citenamefont {Lindsey}\ \emph {et~al.}(2020)\citenamefont
  {Lindsey}, \citenamefont {Goldman}, \citenamefont {Fried},\ and\
  \citenamefont {Bastea}}]{lindsey:jchemphys:2020}%
  \BibitemOpen
  \bibfield  {author} {\bibinfo {author} {\bibfnamefont {R.~K.}\ \bibnamefont
  {Lindsey}}, \bibinfo {author} {\bibfnamefont {N.}~\bibnamefont {Goldman}},
  \bibinfo {author} {\bibfnamefont {L.~E.}\ \bibnamefont {Fried}},\ and\
  \bibinfo {author} {\bibfnamefont {S.}~\bibnamefont {Bastea}},\ }\bibfield
  {title} {\enquote {\bibinfo {title} {Many-body reactive force field
  development for carbon condensation in c/o systems under extreme
  conditions},}\ }\href {https://doi.org/10.1063/5.0012840} {\bibfield
  {journal} {\bibinfo  {journal} {The Journal of Chemical Physics}\ }\textbf
  {\bibinfo {volume} {153}},\ \bibinfo {pages} {054103} (\bibinfo {year}
  {2020})}\BibitemShut {NoStop}%
\bibitem [{\citenamefont {Lindsey}\ \emph {et~al.}(2023)\citenamefont
  {Lindsey}, \citenamefont {Bastea}, \citenamefont {Lyu}, \citenamefont
  {Hamel}, \citenamefont {Goldman},\ and\ \citenamefont
  {Fried}}]{lindsey:jchemphys:2023}%
  \BibitemOpen
  \bibfield  {author} {\bibinfo {author} {\bibfnamefont {R.~K.}\ \bibnamefont
  {Lindsey}}, \bibinfo {author} {\bibfnamefont {S.}~\bibnamefont {Bastea}},
  \bibinfo {author} {\bibfnamefont {Y.}~\bibnamefont {Lyu}}, \bibinfo {author}
  {\bibfnamefont {S.}~\bibnamefont {Hamel}}, \bibinfo {author} {\bibfnamefont
  {N.}~\bibnamefont {Goldman}},\ and\ \bibinfo {author} {\bibfnamefont {L.~E.}\
  \bibnamefont {Fried}},\ }\bibfield  {title} {\enquote {\bibinfo {title}
  {Chemical evolution in nitrogen shocked beyond the molecular stability
  limit},}\ }\href {https://doi.org/10.1063/5.0157238} {\bibfield  {journal}
  {\bibinfo  {journal} {The Journal of Chemical Physics}\ }\textbf {\bibinfo
  {volume} {159}},\ \bibinfo {pages} {084502} (\bibinfo {year}
  {2023})}\BibitemShut {NoStop}%
\bibitem [{\citenamefont {Nance}, \citenamefont {Jakubikova},\ and\
  \citenamefont {Kelley}(2014)}]{nance:jchemtheroycomp:2014}%
  \BibitemOpen
  \bibfield  {author} {\bibinfo {author} {\bibfnamefont {J.}~\bibnamefont
  {Nance}}, \bibinfo {author} {\bibfnamefont {E.}~\bibnamefont {Jakubikova}},\
  and\ \bibinfo {author} {\bibfnamefont {C.~T.}\ \bibnamefont {Kelley}},\
  }\bibfield  {title} {\enquote {\bibinfo {title} {Reaction path following with
  sparse interpolation},}\ }\href {https://doi.org/10.1021/ct5004669}
  {\bibfield  {journal} {\bibinfo  {journal} {Journal of Chemical Theory and
  Computation}\ }\textbf {\bibinfo {volume} {10}},\ \bibinfo {pages}
  {2942--2949} (\bibinfo {year} {2014})}\BibitemShut {NoStop}%
\bibitem [{\citenamefont {Nance}\ and\ \citenamefont
  {Kelley}(2015)}]{nance:jscicomp:2015}%
  \BibitemOpen
  \bibfield  {author} {\bibinfo {author} {\bibfnamefont {J.}~\bibnamefont
  {Nance}}\ and\ \bibinfo {author} {\bibfnamefont {C.~T.}\ \bibnamefont
  {Kelley}},\ }\bibfield  {title} {\enquote {\bibinfo {title} {A sparse
  interpolation algorithm for dynamical simulations in computational
  chemistry},}\ }\href {https://doi.org/10.1137/140965284} {\bibfield
  {journal} {\bibinfo  {journal} {SIAM Journal on Scientific Computing}\
  }\textbf {\bibinfo {volume} {37}},\ \bibinfo {pages} {S137--S156} (\bibinfo
  {year} {2015})}\BibitemShut {NoStop}%
\bibitem [{\citenamefont {Morrow}\ \emph {et~al.}(2021)\citenamefont {Morrow},
  \citenamefont {Kwon}, \citenamefont {Kelley},\ and\ \citenamefont
  {Jakubikova}}]{morrow:jchemtheorycomp:2021}%
  \BibitemOpen
  \bibfield  {author} {\bibinfo {author} {\bibfnamefont {Z.}~\bibnamefont
  {Morrow}}, \bibinfo {author} {\bibfnamefont {H.-Y.}\ \bibnamefont {Kwon}},
  \bibinfo {author} {\bibfnamefont {C.~T.}\ \bibnamefont {Kelley}},\ and\
  \bibinfo {author} {\bibfnamefont {E.}~\bibnamefont {Jakubikova}},\ }\bibfield
   {title} {\enquote {\bibinfo {title} {Efficient approximation of potential
  energy surfaces with mixed-basis interpolation},}\ }\href
  {https://doi.org/10.1021/acs.jctc.1c00569} {\bibfield  {journal} {\bibinfo
  {journal} {Journal of Chemical Theory and Computation}\ }\textbf {\bibinfo
  {volume} {17}},\ \bibinfo {pages} {5673--5683} (\bibinfo {year}
  {2021})}\BibitemShut {NoStop}%
\bibitem [{\citenamefont {Humphrey}, \citenamefont {Dalke},\ and\ \citenamefont
  {Schulten}(1996)}]{humphrey:jmolgrp:1996}%
  \BibitemOpen
  \bibfield  {author} {\bibinfo {author} {\bibfnamefont {W.}~\bibnamefont
  {Humphrey}}, \bibinfo {author} {\bibfnamefont {A.}~\bibnamefont {Dalke}},\
  and\ \bibinfo {author} {\bibfnamefont {K.}~\bibnamefont {Schulten}},\
  }\bibfield  {title} {\enquote {\bibinfo {title} {Vmd: Visual molecular
  dynamics},}\ }\href
  {https://doi.org/https://doi.org/10.1016/0263-7855(96)00018-5} {\bibfield
  {journal} {\bibinfo  {journal} {Journal of Molecular Graphics}\ }\textbf
  {\bibinfo {volume} {14}},\ \bibinfo {pages} {33--38} (\bibinfo {year}
  {1996})}\BibitemShut {NoStop}%
\bibitem [{\citenamefont {Pearce}\ \emph {et~al.}(2021)\citenamefont {Pearce},
  \citenamefont {Wilks}, \citenamefont {Arno},\ and\ \citenamefont
  {O'Reilly}}]{pearce:nature:2021}%
  \BibitemOpen
  \bibfield  {author} {\bibinfo {author} {\bibfnamefont {A.~K.}\ \bibnamefont
  {Pearce}}, \bibinfo {author} {\bibfnamefont {T.~R.}\ \bibnamefont {Wilks}},
  \bibinfo {author} {\bibfnamefont {M.~C.}\ \bibnamefont {Arno}},\ and\
  \bibinfo {author} {\bibfnamefont {R.~K.}\ \bibnamefont {O'Reilly}},\
  }\bibfield  {title} {\enquote {\bibinfo {title} {Synthesis and applications
  of anisotropic nanoparticles with precisely defined dimensions},}\ }\href
  {https://doi.org/10.1038/s41570-020-00232-7} {\bibfield  {journal} {\bibinfo
  {journal} {Nature Reviews Chemistry}\ }\textbf {\bibinfo {volume} {5}},\
  \bibinfo {pages} {21--45} (\bibinfo {year} {2021})}\BibitemShut {NoStop}%
\bibitem [{\citenamefont {Damasceno}, \citenamefont {Engel},\ and\
  \citenamefont {Glotzer}(2012)}]{damasceno:science:2012}%
  \BibitemOpen
  \bibfield  {author} {\bibinfo {author} {\bibfnamefont {P.~F.}\ \bibnamefont
  {Damasceno}}, \bibinfo {author} {\bibfnamefont {M.}~\bibnamefont {Engel}},\
  and\ \bibinfo {author} {\bibfnamefont {S.~C.}\ \bibnamefont {Glotzer}},\
  }\bibfield  {title} {\enquote {\bibinfo {title} {Predictive self-assembly of
  polyhedra into complex structures},}\ }\href
  {https://doi.org/10.1126/science.1220869} {\bibfield  {journal} {\bibinfo
  {journal} {Science}\ }\textbf {\bibinfo {volume} {337}},\ \bibinfo {pages}
  {453--457} (\bibinfo {year} {2012})}\BibitemShut {NoStop}%
\bibitem [{\citenamefont {Nolze}(2015)}]{nolze:cryst:2015}%
  \BibitemOpen
  \bibfield  {author} {\bibinfo {author} {\bibfnamefont {G.}~\bibnamefont
  {Nolze}},\ }\bibfield  {title} {\enquote {\bibinfo {title} {Euler angles and
  crystal symmetry},}\ }\href
  {https://doi.org/https://doi.org/10.1002/crat.201400427} {\bibfield
  {journal} {\bibinfo  {journal} {Crystal Research and Technology}\ }\textbf
  {\bibinfo {volume} {50}},\ \bibinfo {pages} {188--201} (\bibinfo {year}
  {2015})}\BibitemShut {NoStop}%
\bibitem [{\citenamefont {Morrow}\ \emph {et~al.}(2019)\citenamefont {Morrow},
  \citenamefont {Liu}, \citenamefont {Kelley},\ and\ \citenamefont
  {Jakubikova}}]{morrow:jphyschemb:2019}%
  \BibitemOpen
  \bibfield  {author} {\bibinfo {author} {\bibfnamefont {Z.}~\bibnamefont
  {Morrow}}, \bibinfo {author} {\bibfnamefont {C.}~\bibnamefont {Liu}},
  \bibinfo {author} {\bibfnamefont {C.~T.}\ \bibnamefont {Kelley}},\ and\
  \bibinfo {author} {\bibfnamefont {E.}~\bibnamefont {Jakubikova}},\ }\bibfield
   {title} {\enquote {\bibinfo {title} {Approximating periodic potential energy
  surfaces with sparse trigonometric interpolation},}\ }\href
  {https://doi.org/10.1021/acs.jpcb.9b08210} {\bibfield  {journal} {\bibinfo
  {journal} {The Journal of Physical Chemistry B}\ }\textbf {\bibinfo {volume}
  {123}},\ \bibinfo {pages} {9677--9684} (\bibinfo {year} {2019})}\BibitemShut
  {NoStop}%
\bibitem [{\citenamefont {Morrow}\ and\ \citenamefont
  {Stoyanov}(2020)}]{morrow:jscicomp:2020}%
  \BibitemOpen
  \bibfield  {author} {\bibinfo {author} {\bibfnamefont {Z.}~\bibnamefont
  {Morrow}}\ and\ \bibinfo {author} {\bibfnamefont {M.}~\bibnamefont
  {Stoyanov}},\ }\bibfield  {title} {\enquote {\bibinfo {title} {A method for
  dimensionally adaptive sparse trigonometric interpolation of periodic
  functions},}\ }\href {https://doi.org/10.1137/19M1283483} {\bibfield
  {journal} {\bibinfo  {journal} {SIAM Journal on Scientific Computing}\
  }\textbf {\bibinfo {volume} {42}},\ \bibinfo {pages} {A2436--A2460} (\bibinfo
  {year} {2020})}\BibitemShut {NoStop}%
\bibitem [{\citenamefont {Smolyak}(1963)}]{smolyak:dokl:1963}%
  \BibitemOpen
  \bibfield  {author} {\bibinfo {author} {\bibfnamefont {S.}~\bibnamefont
  {Smolyak}},\ }\bibfield  {title} {\enquote {\bibinfo {title} {Quadrature and
  interpolation formulas for tensor products of certain classes of
  functions.}}\ }\href@noop {} {\bibfield  {journal} {\bibinfo  {journal}
  {Dokl. Akad. Nauk SSSR}\ }\textbf {\bibinfo {volume} {148}},\ \bibinfo
  {pages} {1042--1045} (\bibinfo {year} {1963})}\BibitemShut {NoStop}%
\bibitem [{\citenamefont {Judd}\ \emph {et~al.}(2014)\citenamefont {Judd},
  \citenamefont {Maliar}, \citenamefont {Maliar},\ and\ \citenamefont
  {Valero}}]{judd:jecondyncon:2014}%
  \BibitemOpen
  \bibfield  {author} {\bibinfo {author} {\bibfnamefont {K.~L.}\ \bibnamefont
  {Judd}}, \bibinfo {author} {\bibfnamefont {L.}~\bibnamefont {Maliar}},
  \bibinfo {author} {\bibfnamefont {S.}~\bibnamefont {Maliar}},\ and\ \bibinfo
  {author} {\bibfnamefont {R.}~\bibnamefont {Valero}},\ }\bibfield  {title}
  {\enquote {\bibinfo {title} {Smolyak method for solving dynamic economic
  models: Lagrange interpolation, anisotropic grid and adaptive domain},}\
  }\href {https://doi.org/https://doi.org/10.1016/j.jedc.2014.03.003}
  {\bibfield  {journal} {\bibinfo  {journal} {Journal of Economic Dynamics and
  Control}\ }\textbf {\bibinfo {volume} {44}},\ \bibinfo {pages} {92--123}
  (\bibinfo {year} {2014})}\BibitemShut {NoStop}%
\bibitem [{\citenamefont {Petix}\ \emph {et~al.}(2024)\citenamefont {Petix},
  \citenamefont {Fakhraei}, \citenamefont {Kieslich},\ and\ \citenamefont
  {Howard}}]{petix:jchemtheorycomp:2024}%
  \BibitemOpen
  \bibfield  {author} {\bibinfo {author} {\bibfnamefont {C.~L.}\ \bibnamefont
  {Petix}}, \bibinfo {author} {\bibfnamefont {M.}~\bibnamefont {Fakhraei}},
  \bibinfo {author} {\bibfnamefont {C.~A.}\ \bibnamefont {Kieslich}},\ and\
  \bibinfo {author} {\bibfnamefont {M.~P.}\ \bibnamefont {Howard}},\ }\bibfield
   {title} {\enquote {\bibinfo {title} {Surrogate modeling of the relative
  entropy for inverse design using smolyak sparse grids},}\ }\href
  {https://doi.org/10.1021/acs.jctc.3c00651} {\bibfield  {journal} {\bibinfo
  {journal} {Journal of Chemical Theory and Computation}\ }\textbf {\bibinfo
  {volume} {20}},\ \bibinfo {pages} {1538--1546} (\bibinfo {year}
  {2024})}\BibitemShut {NoStop}%
\bibitem [{\citenamefont {Kieslich}, \citenamefont {Boukouvala},\ and\
  \citenamefont {Floudas}(2018)}]{kieslich:jgo:2018}%
  \BibitemOpen
  \bibfield  {author} {\bibinfo {author} {\bibfnamefont {C.~A.}\ \bibnamefont
  {Kieslich}}, \bibinfo {author} {\bibfnamefont {F.}~\bibnamefont
  {Boukouvala}},\ and\ \bibinfo {author} {\bibfnamefont {C.~A.}\ \bibnamefont
  {Floudas}},\ }\bibfield  {title} {\enquote {\bibinfo {title} {Optimization of
  black-box problems using smolyak grids and polynomial approximations},}\
  }\href {https://doi.org/10.1007/s10898-018-0643-0} {\bibfield  {journal}
  {\bibinfo  {journal} {Journal of Global Optimization}\ }\textbf {\bibinfo
  {volume} {71}},\ \bibinfo {pages} {845–869} (\bibinfo {year}
  {2018})}\BibitemShut {NoStop}%
\end{thebibliography}%

\end{document}

\title{Supporting Information: Approximation of anisotropic pair potentials using multivariate interpolation}

\author{Mohammadreza Fakhraei}
\affiliation{Department of Chemical Engineering, Auburn University, Auburn, AL 36849, United States}

\author{Chris A. Kieslich}
\email{kieslich@gatech.edu}
\affiliation{Department of Chemical Engineering, Auburn University, Auburn, AL 36849, United States}
\affiliation{Wallace H. Coulter Department of Biomedical Engineering, Georgia Institute of Technology, Atlanta, Georgia 30332, United States}

\author{Michael P. Howard}
\email{mphoward@auburn.edu}
\affiliation{Department of Chemical Engineering, Auburn University, Auburn, AL 36849, United States}

\maketitle

\section{Symmetry}
We applied symmetry to reduce the relative translational coordinates $\theta$ and $\phi$ as shown in Table 1 using the following procedures:
\begin{itemize}
\item \textit{Rod (2D).} If $\theta > \pi$, rotate by $\pi$ about $z$-axis so $\theta \leq \pi$. Then, if $\theta > \pi/2$, rotate by $\pi$ about $y$-axis so $\theta \leq \pi/2$.
\item \textit{Square.} If $\theta > \pi/2$, rotate by $-\pi/2$ about the $z$-axis until $\theta \leq \pi/2$. Then, if $\theta > \pi/4$, rotate by $\pi$ about the vector $[1/\sqrt{2}, 1/\sqrt{2}]$ so $\theta \leq \pi/4$.
\item \textit{Triangle.} If $\theta > 2\pi/3$, rotate about the $z$-axis by $-2\pi/3$ until $\theta < 2\pi/3$. Then, if $\theta > \pi/3$, rotate by $\pi$ about the vector $[1/2, \sqrt{3}/2]$ so $\theta \leq \pi/3$.
\item \textit{Rod (3D).} If $\phi > \pi/2$, rotate by $\pi$ about the $x$-axis so $\phi \le \pi/2$. Then, rotate by $-\theta$ about the $z$-axis so $\theta = 0$.
\item \textit{Cube.} If $\phi > \pi/2$, rotate by $\pi$ about the $x$-axis so $\phi \le \pi/2$. Next, if $\theta > \pi/2$, rotate about the $z$-axis by $-\pi/2$ until $\theta < \pi/2$. Last, if $\theta > \pi/4$, rotate by $2\pi/3$ about the vector $[1/\sqrt{3}, 1/\sqrt{3}, 1/\sqrt{3}]$ until $\theta \leq \pi/4$.
\item \textit{Tetrahedron.} If $\theta > 2\pi/3$, rotate by $-2\pi/3$ about the $z$-axis until $\theta < 2\pi/3$.
\end{itemize}

\section{Selecting univariate approximations}
\begin{figure}[!h]
    \includegraphics{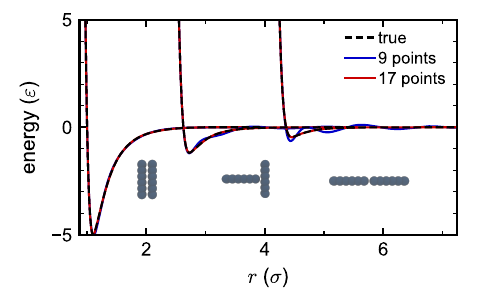}
    \caption{Energy for a pair of two-dimensional rods as a function of $r$ at different fixed $\theta$ and $\alpha$. The three configurations are $(\theta,\alpha) = (\pi/2,0)$, $(0,\pi/2)$, and $(0,0)$, illustrated below the curves. The true energy is compared to Chebyshev polynomial interpolation with respect to $\rho$ using either 9 or 17 sample points. Note the oscillations for the interpolant using 9 sample points for the end-to-end configuration.}
    \label{fig:1d_interp_r}
\end{figure}

\begin{table}[!h]
    \caption{Minimum (min.) and maximum (max.) numbers of sample points considered for each coordinate for $\hat u$.}
    \begin{tabular}{ccccc}
    nanoparticle & \multicolumn{2}{c}{$\rho$} & \multicolumn{2}{c}{angles} \\
     & min. & max. & min. & max. \\
    \hline
    rod (2D), square, triangle   & 17 & 257 & 2 & 33 \\
    rod (3D)  & 17 & 65 & 3 & 65\\
    cube, tetrahedron  & 17 & 33 & 3 & 33 
    \end{tabular}
    \label{tab:min_max_points}
\end{table}

\begin{table}[!h]
    \caption{Number of sample points for each coordinate for the best linear piecewise (L), Chebyshev polynomial (C), and mixed-basis polynomial (M) interpolants found for $\hat u$.}
    \begin{tabular}{cccccccc}
    nanoparticle & interpolant & $\rho$ & $\theta$ & $\phi$ & $\alpha$ & $\beta$ & $\gamma$\\
    \hline
    \multirow{3}{*}{rod (2D)}  & L & 33 & 5 & & 9 & & \\
    & C & 17 & 9 & & 9 & & \\
    & M & 17 & 9 & & 9 & & \\
    \hline
    \multirow{3}{*}{square}  & L & 65 & 5 & & 5 & & \\
    & C & 17 & 9 & & 9 & & \\
    & M & 17 & 9 & & 9 & & \\
    \hline
    \multirow{3}{*}{triangle}  & L & 33 & 5 & & 9 & & \\
    & C & 17 & 5 & & 17 & & \\
    & M & 17 & 9 & & 9 & & \\
    \hline
    \multirow{3}{*}{rod (3D)}  & L & 17 & & 5 & 17 & 5 & \\
    & C & 17 & & 5 & 17 & 5 & \\
    & M & 17 & & 9 & 9 & 5 & \\
    \hline
    \multirow{3}{*}{cube}  & L & 33 & 3 & 3 & 17 & 3 & 3 \\
    & C & 17 & 3 & 5 & 17 & 3 & 3 \\
    & M & 17 & 3 & 9 & 9 & 3 & 3 \\
    \hline
    \multirow{3}{*}{tetrahedron}  & L & 33 & 3 & 5 & 5 & 5 & 3 \\
    & C & 17 & 3 & 9 & 3 & 5 & 5 \\
    & M & 17 & 3 & 17 & 3 & 5 & 3
    \end{tabular}
    \label{tab:points_combination}
\end{table}

\begin{table}[!h]
    \caption{Number of sample points for each coordinate for Chebyshev polynomial interpolants $\hat r_0$ with varied RMSE.}
    \begin{tabular}{ccccccc}
    nanoparticle & $\theta$ & $\phi$ & $\alpha$ & $\beta$ & $\gamma$ & RMSE ($\sigma$)\\
    \hline
    \multirow{4}{*}{rod (2D)} & 9 & & 9 & & & 0.039 \\
    & 17 & & 33 & & & 0.014 \\
    & 33 & & 65 & & & 0.007 \\
    & 65 & & 128 & & & 0.001 \\
    \hline
    \multirow{4}{*}{rod (3D)} & & 5 & 17 & 5 & & 0.104 \\
    & & 9 & 33 & 5 & & 0.037 \\
    & & 17 & 33 & 9 & & 0.02 \\
    & & 33 & 65 & 17 & & 0.007 \\
    \end{tabular}
    \label{tab:min_max_points_r0}
\end{table}

\clearpage
\section{Two-dimensional nanoparticles}
\begin{figure}[!h]
    \includegraphics{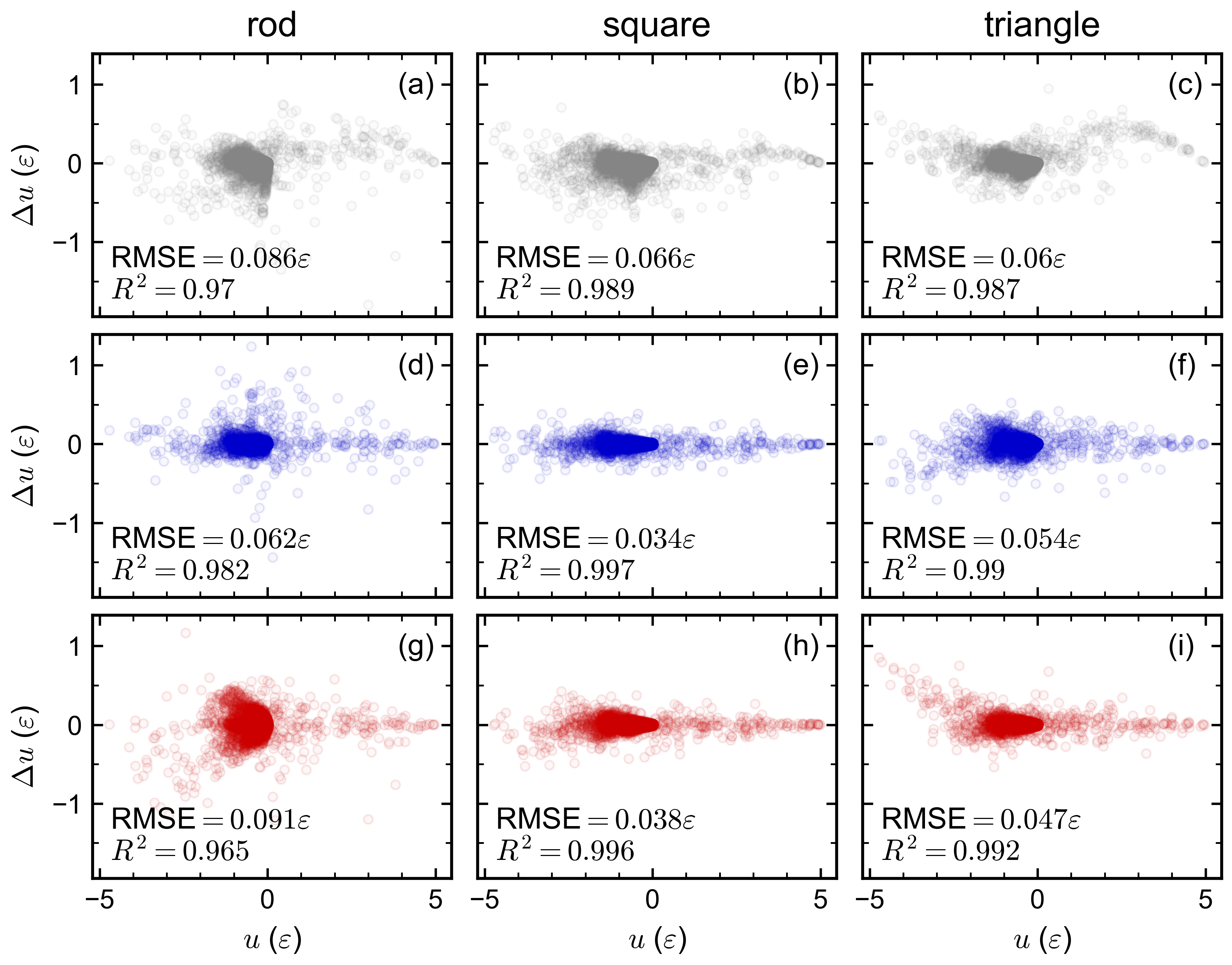}
    \caption{Same as Figure 5, but showing the residual energy $\Delta u = \hat u - u$ vs.~the true energy $u$.}
    \label{fig:residual_2d}
\end{figure}
\begin{figure}[!h]
    \includegraphics{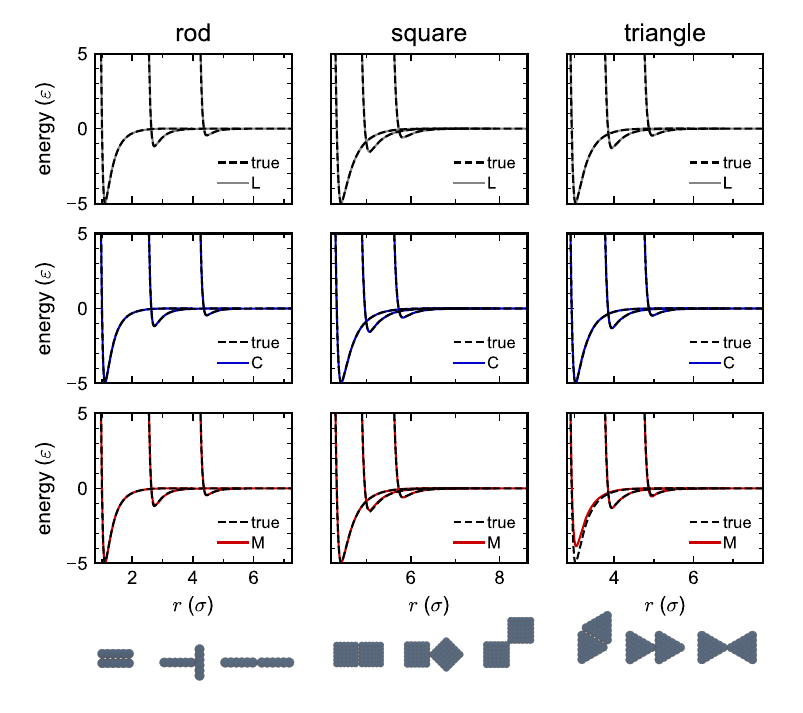}
    \caption{Comparison of energy approximated using linear piecewise (L), Chebyshev polynomial (C), and mixed-basis polynomial (M) interpolants to the true energy for two-dimensional nanoparticles as a function of $r$ at different fixed $\theta$ and $\alpha$. The three configurations $(\theta,\alpha)$ are the same for the rod as in Figure \ref{fig:1d_interp_r}; $(0,0)$, $(0,\pi/4)$, and $(\pi/4,0)$ for the square; and $(\pi/3,\pi/3)$, $(0,0)$, and $(0,\pi/3)$ for the triangle, illustrated at the bottom.}
    \label{fig:u_r_2d}
\end{figure}

\clearpage
\section{Three-dimensional nanoparticles}
\begin{figure}[!h]
    \includegraphics{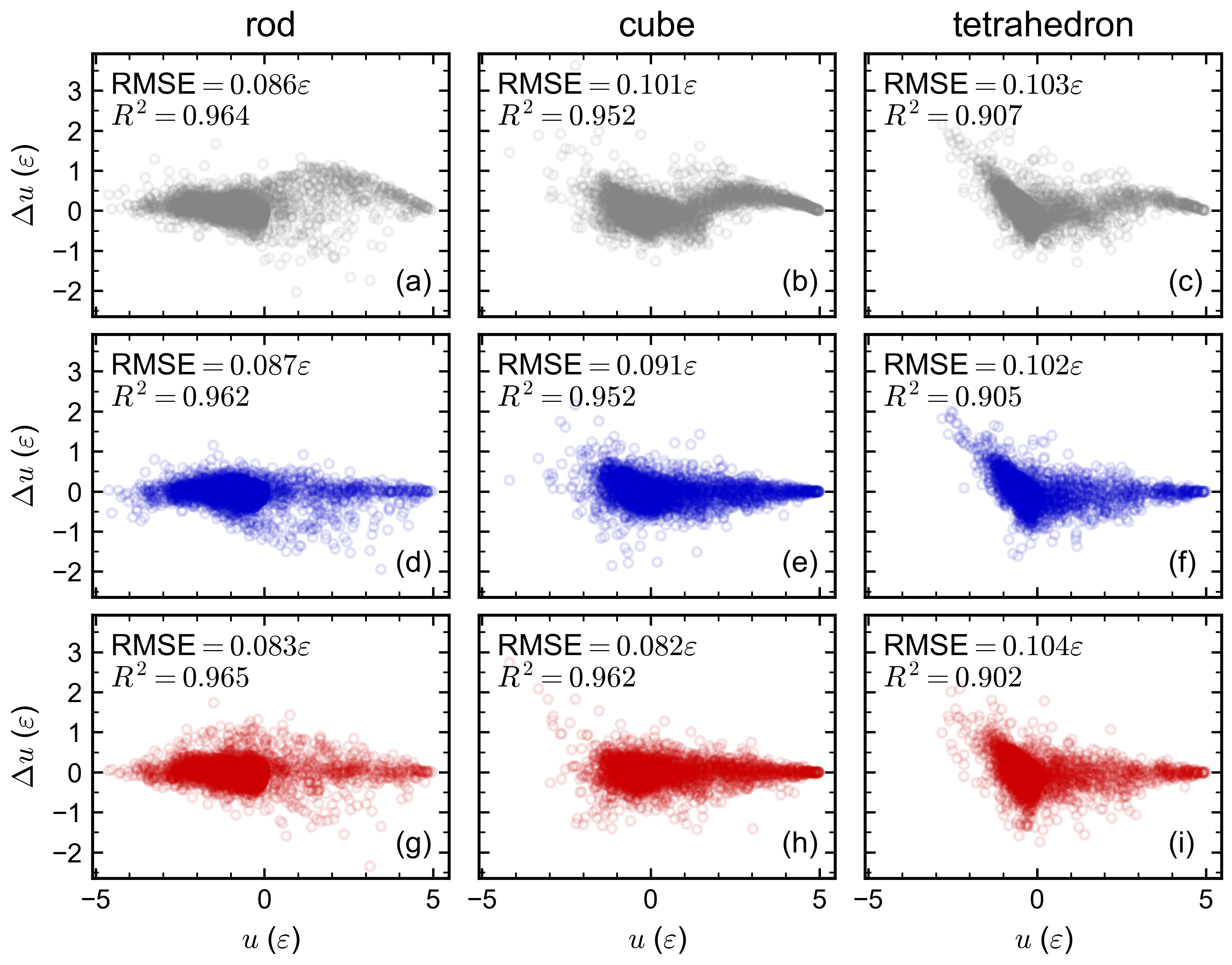}
    \caption{Same as Figure 8, but showing the residual energy $\Delta u = \hat u - u$ vs.~the true energy $u$.}
    \label{fig:residual_3d}
\end{figure}
\begin{figure}[!h]
    \includegraphics{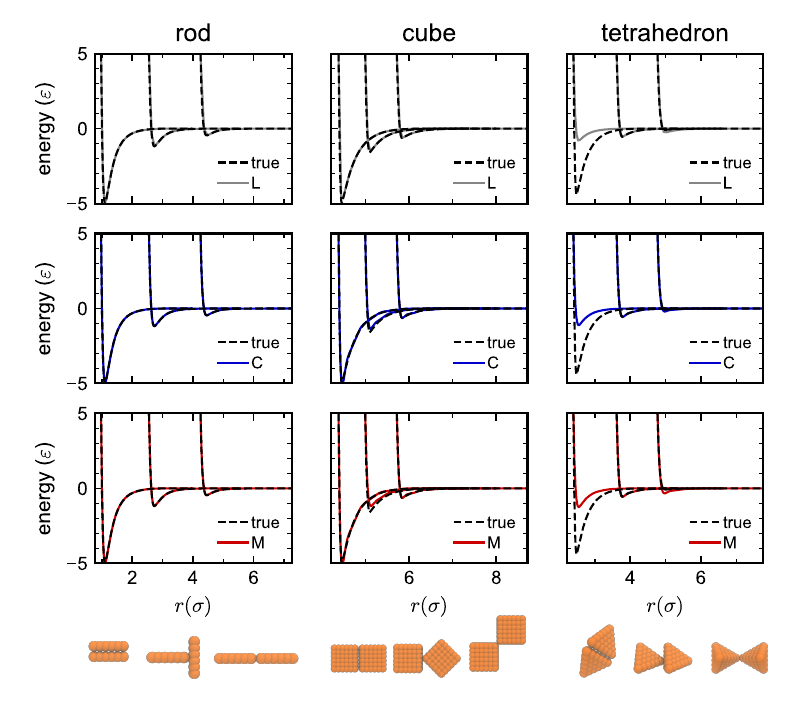}
    \caption{Comparison of energy approximated using linear piecewise (L), Chebyshev polynomial (C), and mixed-basis polynomial (M) interpolants to the true energy for the three-dimensional nanoparticles as a function of $r$ at different fixed $\theta$, $\phi$, and orientation. The three configurations are $(\phi, \alpha, \beta) = (\pi/2,0,0)$, $(0,\pi/2,\pi/2)$, and $(0,0,0)$ for the rod; $(\theta, \phi, \alpha, \beta, \gamma) = (0,\pi/2,0,0,0)$, $(0,\pi/2,\pi/4,0,0)$, and $(\pi/4,\pi/2,0,0,0)$ for the cube; and $(\theta, \phi, \alpha, \beta, \gamma) = (4\pi/3,\pi,\pi/2,7\pi/18,\pi/6)$, $(0,0,0,0,0)$, and $(2\pi/3,\pi/2,\pi/3,0,0)$ for the tetrahedron, illustrated at the bottom.}
    \label{fig:u_r_3d}
\end{figure}